\documentclass[a4paper,11pt]{article}

\usepackage{xcolor}
\usepackage{jheppub}
\pdfoutput=1

\usepackage{mathtools} 
\usepackage{graphicx}
\usepackage{dcolumn}
\usepackage{bm}
\usepackage{slashed}
\usepackage{color}
\usepackage{physics}
\usepackage{comment}
\usepackage[normalem]{ulem}
\usepackage{hyperref}
\usepackage{url}

\usepackage{physics}

\usepackage{tikz}
\usepackage{tikz-3dplot}
\usetikzlibrary{arrows.meta, patterns}
\usetikzlibrary{positioning, shapes.geometric}
\usetikzlibrary{decorations.pathreplacing}

\tikzset{
  pics/torus/.style n args={3}{
    code = {
      \providecolor{pgffillcolor}{rgb}{1,1,1}
      \begin{scope}[
          yscale=cos(#3),
          outer torus/.style = {draw,line width/.expanded={\the\dimexpr2\pgflinewidth+#2*2},line join=round},
          inner torus/.style = {draw=pgffillcolor,line width={#2*2}}
        ]
        \draw[outer torus] circle(#1);\draw[inner torus] circle(#1);
        \draw[outer torus] (180:#1) arc (180:360:#1);\draw[inner torus,line cap=round] (180:#1) arc (180:360:#1);
      \end{scope}
    }
  }
}

\begin{document}
\count\footins = 1000

\preprint{KEK-TH-2761, IPMU25-0048}

\title{\boldmath Magnetic Helicity, Magnetic Monopoles, and Higgs Winding}
\author[a]{Hajime Fukuda,}
\author[b,c]{Yuta Hamada,}
\author[d,e,f]{Kohei Kamada,}
\author[b,c]{Kyohei Mukaida}
\author[g]{and Fumio Uchida}

\affiliation[a]{Department of Physics, The University of Tokyo, Tokyo 113-0033, Japan}
\affiliation[b]{Theory Center, IPNS, KEK, Ibaraki 305-0801, Japan}
\affiliation[c]{Graduate University for Advanced Studies (SOKENDAI), Ibaraki 305-0801, Japan}
\affiliation[d]{School of Fundamental Physics and Mathematical Sciences, Hangzhou Institute for Advanced Study, University of Chinese Academy of Sciences (HIAS-UCAS), 310024 Hangzhou, China}
\affiliation[e]{International Centre for Theoretical Physics Asia-Pacific (ICTP-AP), Hangzhou/Beijing, China}
\affiliation[f]{RESCEU, The University of Tokyo, Bunkyo-ku, Tokyo 113-0033, Japan}

\emailAdd{hfukuda@hep-th.phys.s.u-tokyo.ac.jp}
\emailAdd{yhamada@post.kek.jp}
\emailAdd{kohei.kamada@ucas.ac.cn}
\emailAdd{kyohei.mukaida@kek.jp}
\emailAdd{fumio.uchida@ipmu.jp}
\affiliation[g]{Kavli IPMU (WPI), UTIAS, The University of Tokyo, Kashiwa, Chiba 277-8583, Japan}

\date{\today}

\abstract{
    Changes in magnetic helicity are often discussed across a variety of fields,
    from condensed matter physics to early universe cosmology.
    It is frequently stated that the helicity change is given by the integral of the gauge field strength tensor and its dual over spacetime, $\int F \wedge F$. However, this is incorrect when magnetic monopoles once exist in the spacetime. In this paper, we show the correct formula of the helicity change in such a case for the Maxwell theory with the magnetic monopoles. 
    We also discuss what happens when we embed the Maxwell theory with magnetic monopoles into non-Abelian gauge theories. We show that a similar formula holds for the 't Hooft--Polyakov monopole. In particular, we find the winding numbers and the zeroes of the Higgs field in the non-Abelian gauge theory play a crucial role in the helicity change. 
    The same discussion is also applicable to the electroweak theory, and we discuss the implication of our findings to the baryon number change via the chiral anomaly in the early universe.
}

\maketitle
\newpage

\section{Introduction and summary}
The existence of the intergalactic magnetic field is one of the most intriguing phenomena in modern astrophysics\,\cite{Neronov:2010gir,Tavecchio:2010mk,Dolag:2010ni,Fermi-LAT:2018jdy,MAGIC:2022piy}. The origin of the intergalactic magnetic field is still an open question, but one of the interesting possibilities is that it is generated in the early universe, called the primordial magnetic field\,\cite{Grasso:2000wj,Durrer:2013pga,Subramanian:2015lua}. The primordial magnetic field, once generated, evolves with the thermal plasma in the early universe. The evolution of the primordial magnetic field is described by the coupled dynamics of the electromagnetic field and the plasma, which is called the magnetohydrodynamics (MHD) theory\,\cite{Son:1998my,Jedamzik:1996wp,Christensson:2000sp,Banerjee:2004df,Durrer:2013pga,Subramanian:2015lua,biskamp2003magnetohydrodynamic}. One of the important features of the MHD theory is that the magnetic helicity is well conserved if the magnetic Reynolds number is large enough~\cite{Frisch_Pouquet_LEOrat_Mazure_1975,Banerjee:2004df,Campanelli:2007tc,Kahniashvili:2012uj}.
The conservation of the magnetic helicity is crucial for understanding the evolution of the  
magnetic field --- not only primordial ones in the early universe, but also turbulent MHD systems in general ---  as it provides a constraint on the dynamics of the electromagnetic field and plasma.

The magnetic helicity in some volume $V$ is defined as the integral of the magnetic field and the vector potential\,\cite{1978magnetic}:
\begin{align*}
    \mathcal{H} = \int_V d^3x \, \vec{A} \cdot \vec{B},
\end{align*}
where $\vec{B} = \nabla \times \vec{A}$ is the magnetic field and $\vec{A}$ is the vector potential.
The change of the magnetic helicity is often considered to be equal to the integral of the electromagnetic field strength $F$ over the spacetime manifold, $\int F \wedge F$ as\,\cite{Giovannini:1997eg}:
\begin{align*}
    \Delta \mathcal{H} \stackrel{?}{=} \int F \wedge F.
\end{align*}
Once we accept this relation, the change of the magnetic helicity is related to the change of the chiral charge of fermions via the Adler--Bell--Jackiw (ABJ) anomaly\,\cite{Adler:1969gk,Bell:1969ts}, i.e., $\Delta Q_5 \propto \int F \wedge F \stackrel{?}{=} \Delta \mathcal{H}$, leading to various interesting implications.
The transport phenomena induced by this relation drives distinct dynamics in the system (see e.g., \cite{Boyarsky:2011uy,Hirono:2015rla}).
Similar relation also holds in the axion electrodynamics, where the charge carried by axion plays the same role to the chirality\,\cite{Domcke:2018eki,Choi:2022fgx,Yokokura:2022alv}.
Furthermore, the magnetic helicity in the universe may be related to the baryon asymmetry of the universe through the ABJ anomaly\,\cite{Giovannini:1997gp,Giovannini:1997eg,Fujita:2016igl,Kamada:2016eeb,Kamada:2016cnb,Chao:2024fip,Fukuda:2024pkh}.

The main purpose of this paper is to show that the above relation between the change in magnetic helicity, $\Delta \mathcal{H}$, and $F \wedge F$ does not generally hold, and to provide the correct formula when the magnetic monopoles once exist in the spacetime.
The subtlety arises since the presence of magnetic monopoles obstructs a global definition of the vector potential $\vec{A}$\,\cite{Dirac:1948um,Sakurai:2011zz}, so does the magnetic helicity.
Although observational constraints limit the current monopole abundance\,\cite{Parker:1970xv,Kolb:1982si,Orito:1990ny,MACRO:2002jdv,Super-Kamiokande:2012tld}, they could have existed in the early universe\,\cite{Langacker:1980kd}, e.g., at the stage of magnetogenesis, or been produced via the Schwinger effect in the presence of a strong magnetic field.
In the electroweak theory, an unstable dumbbell-like object, which is a pair of monopole and anti-monopole connected by a $Z$-boson string, are sometimes considered in the electroweak phase transition\,\cite{Nambu:1977ag,Vachaspati:1992fi}, or one may also consider more generic configurations of the Higgs field and gauge fields.
It is therefore of importance to clarify how the existence of monopoles modifies the relation between $\Delta \mathcal{H}$ and $\int F \wedge F$ in order to correctly understand the dynamics of the magnetic field and its implications on the baryon asymmetry of the universe.

\vskip1em
The structure and the summary of this paper are as follows.
In Sec.\,\ref{sec:abelian_helicity_change}, we study the helicity and its change in the effective $\text{U}(1)$ gauge theory with magnetic monopoles and dyons. 
We show that the helicity change formula is modified from the Chern number, $\int F \wedge F$, by the contribution of the monopoles and dyons as
\begin{align*}
    \Delta \mathcal{H} = \int_{M'} F \wedge F + \frac{4\pi}{e} \sum_i \int_{\Sigma_i} F,
\end{align*}
where $M'$ is the spacetime manifold excluding the worldlines of the monopoles and dyons, $\Sigma_i$ is the worldsheet of the Dirac string spanning the $i$-th 't Hooft loop, and $e$ is the gauge coupling.
This formula is the main result of this paper.

In Sec.\,\ref{sec:non_abelian_adjoint_Higgs}, we study the helicity and its change in the Georgi--Glashow model\,\cite{Georgi:1972cj} as an artificial but simple UV completion of the $\text{U}(1)$ gauge theory with monopoles.
We show that the formula similar to the one in Sec.\,\ref{sec:abelian_helicity_change} holds.
We also show that the helicity can be expressed in terms of the UV quantities as
\begin{align*}
    \mathcal{H} = \frac{16 \pi^2}{g^2} \qty( N_{\text{CS}} + N_{H} ),
\end{align*}
where $g$ is the $\text{SU}(2)$ gauge coupling, $N_{\text{CS}}$ is the Chern--Simons number of the $\text{SU}(2)$ gauge field, and $N_H$ is the winding number of the adjoint Higgs field.

In Sec.\,\ref{sec:standard_model_like}, we study the helicity and its change in the electroweak-like gauge theory.
With an appropriate reinterpretation of $\Sigma_i$, 
we find the formula similar to the one in Sec.\,\ref{sec:abelian_helicity_change} holds.
We also show that the helicity can be expressed in terms of $N_{\text{CS}}$ and $N_{H}$ as conjectured in Ref.\,\cite{Hamada:2025cwu}.
This implies that there is no one-to-one correspondence between the helicity and  baryon asymmetry, and hence the previous estimation based on this assumption should be reconsidered.

Finally,
Sec.\,\ref{sec:conclusion} is devoted to the conclusion and discussion. 

\section{Helicity in Maxwell theory with magnetic monopoles}
\label{sec:abelian_helicity_change}
In this section, we examine the concept of helicity in Maxwell theory in the presence of magnetic monopoles. We establish the helicity change formula even when monopoles and dyons once exist in the spacetime manifold. 
Let us first review the helicity and its change in the Maxwell theory without monopoles. The helicity on a time slice at $t_i$ 
is written as
\begin{align}
    \mathcal{H} \equiv \int d^3x \, \vec{A} \cdot \vec{B} \bigg|_{t=t_i} = 
    \frac12 \int d^3x \, \epsilon^{ijk} F_{ij} A_k \bigg|_{t=t_i}. \label{eq:helicity_def_no_monopole}
\end{align}
This is invariant under small gauge transformations as long as the spatial volume is closed or $F_{ij}$ vanishes at the spatial infinity.
The definition of the four vector potential $A_\mu$ and field strength $F_{\mu\nu}$ are given in App.\,\ref{app:notations}.
For the later convenience, we introduce the differential form notation. The helicity on any $3$-dimensional hypersurface $\mathcal{S}$ is written as
\begin{align}
    \mathcal{H} = \int_\mathcal{S} A \wedge F, \label{eq:helicity_def_no_monopole_diff_form}
\end{align}
where $A$ and $F = dA$, defined in App.\,\ref{app:notations}, are the $1$-form vector potential and the $2$-form field strength, respectively. Using the differential form notation, we can easily write the change of the helicity from a past hypersurface $\mathcal{S}_\text{past}$ to a future hypersurface $\mathcal{S}_\text{future}$ as
\begin{align}
    \Delta \mathcal{H} \equiv \mathcal{H}_\text{future} - \mathcal{H}_\text{past} = \int_M d\qty(A \wedge F) = 
     \int_M F \wedge F = -2 \int_M \vec{E} \cdot \vec{B} \, d^4x, \label{eq:helicity_change_no_monopole}
\end{align}
where $M$ is the $4$-dimensional spacetime manifold bounded by $\mathcal{S}_\text{past}$ and $\mathcal{S}_\text{future}$. The main purpose of this section is to see how the above helicity change formula, Eq.\,\eqref{eq:helicity_change_no_monopole}, is modified when monopoles or dyons exist in $M$.

First of all, the magnetic helicity is not invariant under the large gauge transformation, which appears in the presence of magnetic monopoles\,\cite{Dirac:1948um,Sakurai:2011zz}. When a magnetic monopole is present, the vector potential is not globally well-defined. Instead, the vector potential is defined locally. We can typically take two local patches, $U_N$ and $U_S$, covering the north and south hemispheres of the monopole, respectively, when we consider a sphere surrounding the monopole. The vector potential defined on $U_N$ and $U_S$ is related by the large gauge transformation.

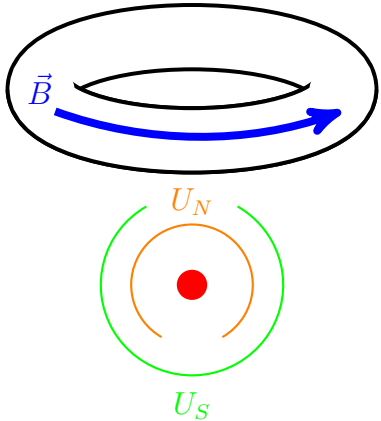
\begin{figure}[t]
\begin{center}
\begin{tikzpicture}
    \pgfsetlinewidth{1.5pt}
 \pic{torus={2cm}{4mm}{70}};
    \draw[blue,line width=3pt,->,>=stealth'] (-1.8cm, -0.3cm) arc [radius=5.5cm, start angle=250, end angle=290];
    \node[blue] at (-2cm, 0) {\large $\vec{B}$};
    \fill[red] (0, -2.6cm) circle (0.2cm);
    \draw[orange,thick] (0, -2.6cm)++(240:0.8cm) arc [radius=0.8cm, start angle=240, end angle=-60];
    \node[orange] at (0, -1.5cm) {\large $U_N$};
    \draw[green,thick] (0, -2.6cm)++(120:1.2cm) arc [radius=1.2cm, start angle=120, end angle=420];
    \node[green] at (0, -4.2cm) {\large $U_S$};
  \end{tikzpicture}
\end{center}
\caption{An example of the configuration where the helicity is not well-defined. Some current supports the constant magnetic field $\vec{B}$ inside the torus. The red blob represents a magnetic monopole. The local patches $U_N$ and $U_S$ are shown in orange and green, respectively. The vector potential is defined in the direction of the orange and green surfaces from the center of the monopole. The two vector potentials are related by the large gauge transformation. 
}
\label{fig:helicity_not_well_defined}
\end{figure}

We show an example of the configuration where the helicity is ill-defined in Fig.\,\ref{fig:helicity_not_well_defined}. A magnetic field is supported inside the torus by a current, and the magnetic monopole is placed below the toroidal flux. Let us consider the helicity in the torus. Assuming the magnetic field and the vector potential are constant inside the torus, the helicity is supposed to be $\mathcal H \propto \oint A_\varphi$, where $A_\varphi$ is the vector potential in the merdian direction of the torus and the integral is taken over the merdian circle of the torus.
If we use the vector potential defined on $U_N$, $A_\varphi$ almost vanishes inside the torus and the helicity is zero. On the other hand, if we use the vector potential defined on $U_S$, $A_\varphi$ is non-vanishing and the helicity is nonzero.
Stating in another way, the integral $\oint A_\varphi$ can be rewritten as the integral of the magnetic field over a surface whose boundary is the merdian circle of the torus, using the Stokes theorem. However, the magnetic flux through the surface is dependent on the choice of the surface; if the surface encloses the monopole, the magnetic flux is nonzero, while if the surface does not enclose the monopole, the magnetic flux is almost zero. 

Therefore, to discuss the helicity, we need to assume that all monopoles and dyons are, even if they once exist, created after the hypersurface $\mathcal{S}_\text{past}$ and annihilated before the hypersurface $\mathcal{S}_\text{future}$. Assuming that the monopoles and dyons are much heavier than the energy scale of the interest, we can treat them as external classical objects in the spacetime manifold $M$.
The existence of such particles in the spacetime manifold is described by $1$-dimensional line, the \emph{worldline} of the monopole or dyon. In particular, they are pairwise created and annihilated, so the worldlines of the monopoles and dyons are closed loops, which we call 't Hooft loops and Wilson--'t Hooft loops, respectively. 

\begin{figure}[t]
\begin{center}
    \begin{minipage}{0.45\hsize}
        \begin{tikzpicture}
        \node[anchor=south west, inner sep=0] at (0,0) {\includegraphics[width=\textwidth]{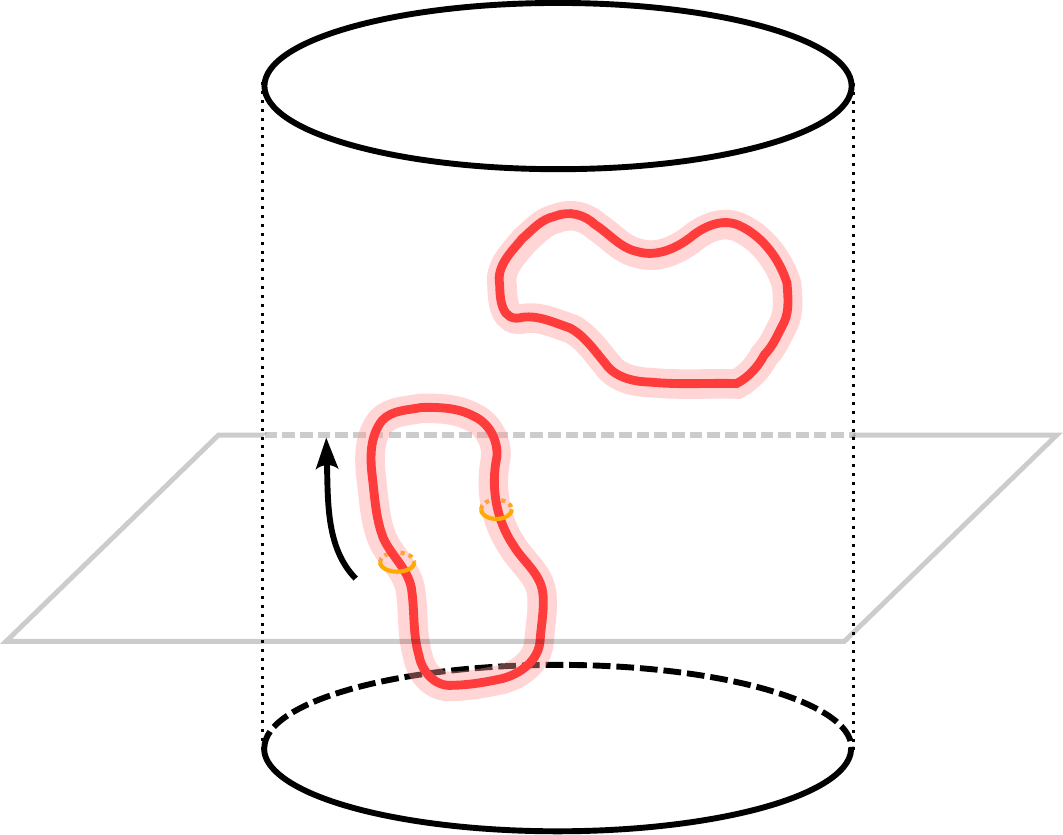}};
        \node[red] at (2.7cm, 3.05cm) {\large $L_i$};
        \node[red] at (5.2cm, 4.0cm) {\large $L_j$};
        \node[black] at (3.65cm, 0.55cm) {Past hypersurface};
        \node[black] at (3.65cm, 4.8cm) {Future hypersurface};
        \node[gray, xslant=1] at (1.2cm, 1.5cm) {timeslice};
        \end{tikzpicture}
    \end{minipage}
    \hspace{0.05\hsize}
    \begin{minipage}{0.45\hsize}
        \begin{tikzpicture}
        \node[anchor=south west, inner sep=0] at (0,0) {\includegraphics[width=0.95\textwidth]{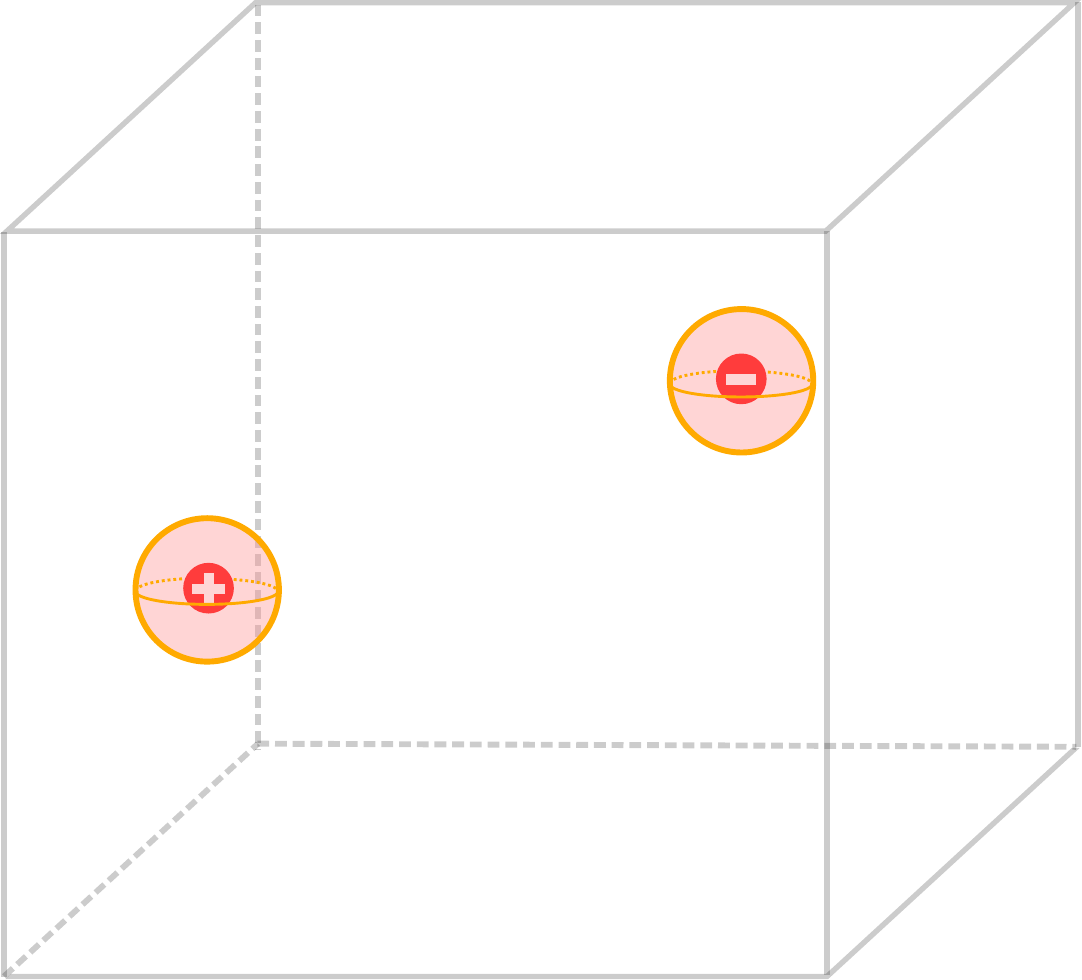}};
        \end{tikzpicture}
    \end{minipage}
\end{center}
\caption{(Left)
A schematic representation of the punctured manifold $M'$ with neighbourhoods of 't Hooft loops.
't Hooft loops $L_i$ and $L_j$ are shown as red loops. The scooped-out regions around the loops are shown as pink regions.
        The arrow near $L_i$ indicates the direction of the 't Hooft loop.
        (Right) A schematic picture when we take a time slice of the punctured manifold $M'$. The time slice is shown in the left figure as a gray slanted rectangle. The 't Hooft line is a pair of monopole and anti-monopole punctures on the time slice, which are shown as red dots with $+$ and $-$ signs, respectively. 
        The intersection of the time slice with $L_i$ is shown as an orange spherical surface. The scooped regions are shown as pink balls surrounding the monopole and anti-monopole punctures.
        }
\label{fig:thooft_loop_cut_off}
\end{figure}
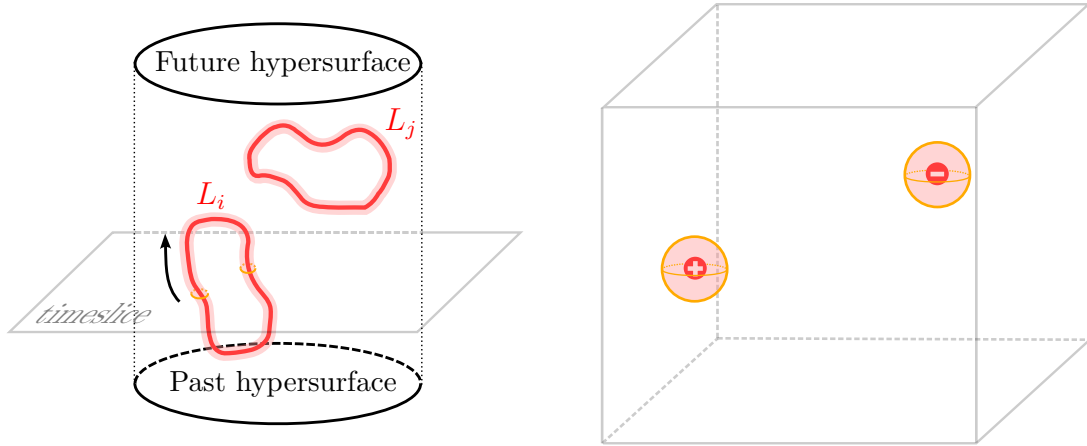

For now, let us consider only the magnetic monopoles of unit charges once exist, i.e., only elementary 't Hooft loops are present in the spacetime manifold $M$. We also assume that the spacetime manifold $M$ does not have a boundary except for the initial and final time slices at this point.
Just as point-like electric charges introduce singularities into the Maxwell theory, magnetic monopoles introduce singularities as well; the electromagnetic fields are well-defined only on a point outside the worldlines of the monopoles\,\footnote{Do not confuse this with the fact that, unlike the electric charges, the vector potential is not well-defined globally, even outside the monopole worldlines.}. 
Therefore, \emph{to define the Maxwell theory},
we need to remove the neighbourhood of the worldlines of the magnetic monopoles from $M$.
The neighbourhood of each monopole worldline is denoted by $L_i = \mathcal{C}_i \times B^3$, where $i$ labels the $i$-th monopole. $\mathcal{C}_i \cong S^1$ corresponds to the worldline of the monopole, and $B^3$ is a three-dimensional ball surrounding the monopole if we consider a time slice. The radius of the ball is the inverse of the UV cutoff scale, which is taken to be much smaller than any length scale of the interest.
We also require that these punctures are mutually disjoint: $L_i \cap L_j = \emptyset$ for $i \neq j$.
The resulting punctured manifold is $M' = M \setminus \bigcup_i L_i$.
For simplicity, we neglect point-like electric charges in this discussion; if present, their worldlines (Wilson loops) would also need to be excised from the manifold, or we would need to regularize the singularity by smearing the electric charge over a small region. We show the schematic picture of the punctured manifold $M'$ in Fig.\,\ref{fig:thooft_loop_cut_off}.

On the punctured manifold $M'$, Maxwell theory is well-defined, and the electromagnetic field strength $F_{\mu\nu}$ can be specified. However, defining the helicity, $\int d^3x \, \epsilon^{ijk} F_{ij} A_k$, requires a well-defined vector potential $A_\mu$. In general, $A_\mu$ cannot be defined globally on $M'$ because the Chern number $\int F$ over a closed surface surrounding a puncture is nonzero\,\cite{Nakahara:2003nw}; integrating $F_{\mu\nu}$ over an $S^2$ enclosing a monopole yields its magnetic charge\,\footnote{Note that in four dimensions, two-dimensional surfaces and one-dimensional loops can be linked.}. Stating differently, the vector potential $A_\mu$ is not globally well-defined on $M'$, since monopoles once exist in the spacetime manifold $M'$.

However, on the past and future hypersurfaces, as no monopole is present, there is a natural definition of the vector potential $A_\mu$ up to small gauge transformations. Since we are interested in the change of the helicity, we would like to connect these two vector potentials smoothly. It is possible if we separate the spacetime manifold $M'$ into several patches. For each patch, we can define the vector potential $A_\mu$ globally inside the patch, if the Chern number is vanishing in the patch. If one of the patches contains both the past and future hypersurfaces, we have the vector potential $A_\mu$ defined on both hypersurfaces up to small gauge transformations and therefore the helicity change is well-defined.

Then, how can we take these patches? We would like to take a region which contains both the past and future hypersurfaces, but does not support any nontrivial Chern number, i.e., the region does not contain any surface that encloses a monopole or anti-monopole puncture.
To achieve this, for each monopole worldline neighbourhood $L_i$, we enclose it within a closed three-dimensional surface $S_i \cong S^3$. Let $B_i$ denote the four-dimensional region bounded by $S_i$ and containing $L_i$, with all $B_i$ chosen to be mutually disjoint for the time being. The specific choice of $B_i$ is not unique, but the final physical results will not depend on this choice, as discussed later. 
We now define a patch, or a new manifold $M'' = M \setminus \bigcup_i B_i$, where all regions supporting nontrivial Chern numbers have been excised. On $M''$, the vector potential $A_\mu$ can be defined globally, which connect the past and future hypersurfaces.

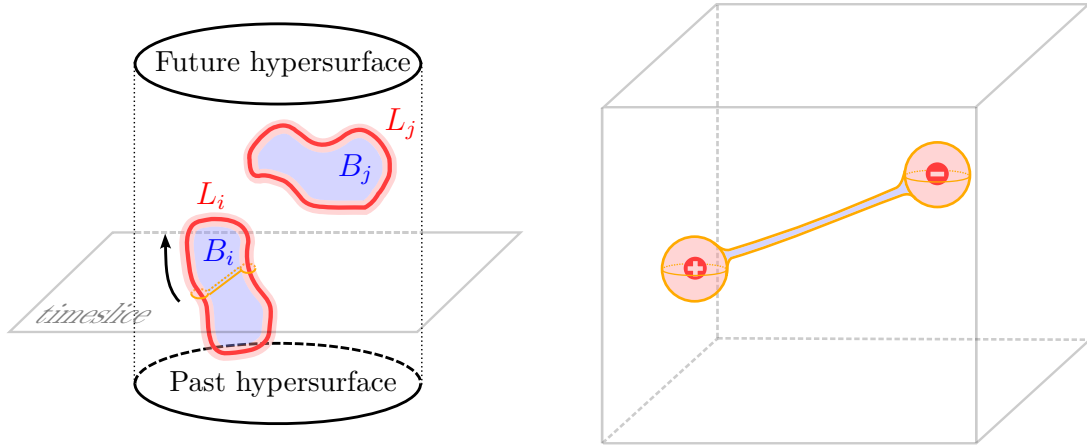
\begin{figure}[t]
\begin{center}
    \begin{minipage}{0.45\hsize}
        \begin{tikzpicture}
        \node[anchor=south west, inner sep=0] at (0,0) {\includegraphics[width=\textwidth]{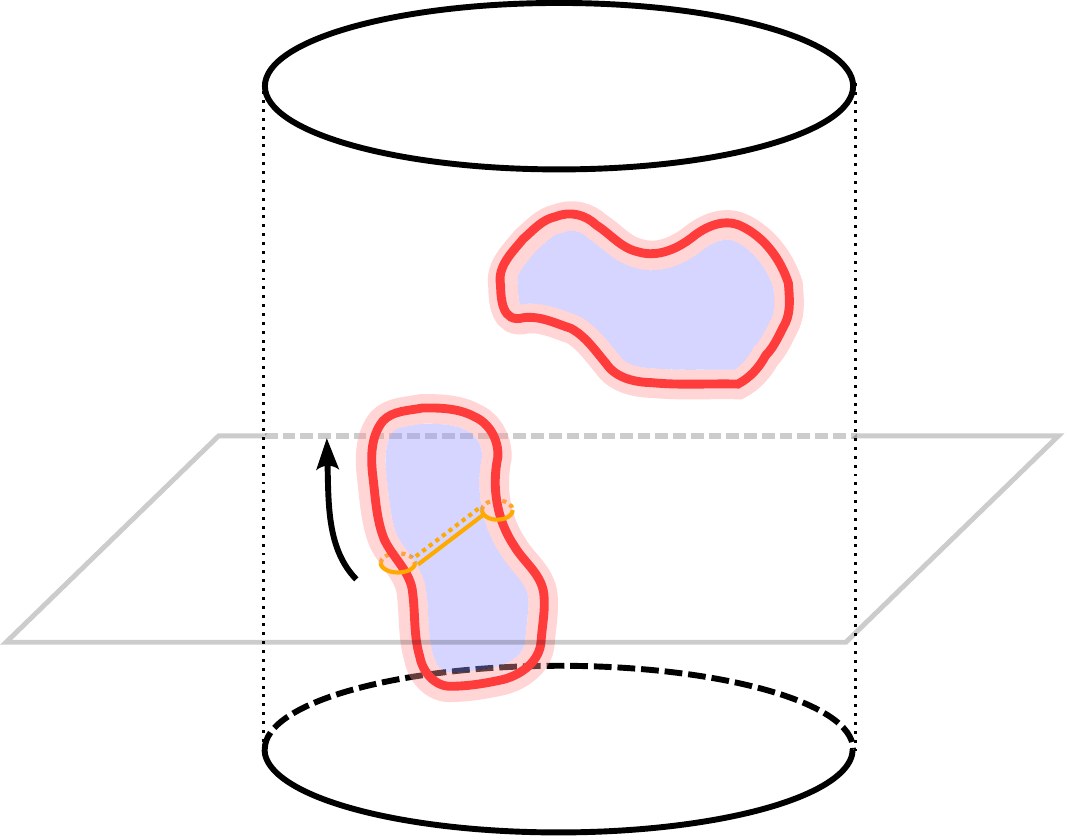}};
        \node[blue] at (2.8cm, 2.3cm) {\large $B_i$};
        \node[blue] at (4.6cm, 3.4cm) {\large $B_j$};

        \node[red] at (2.7cm, 3.05cm) {\large $L_i$};
        \node[red] at (5.2cm, 4.0cm) {\large $L_j$};
        \node[black] at (3.65cm, 0.55cm) {Past hypersurface};
        \node[black] at (3.65cm, 4.8cm) {Future hypersurface};
        \node[gray, xslant=1] at (1.2cm, 1.5cm) {timeslice};
        \end{tikzpicture}
    \end{minipage}
    \hspace{0.05\hsize}
    \begin{minipage}{0.45\hsize}
        \begin{tikzpicture}
        \node[anchor=south west, inner sep=0] at (0,0) {\includegraphics[width=0.95\textwidth]{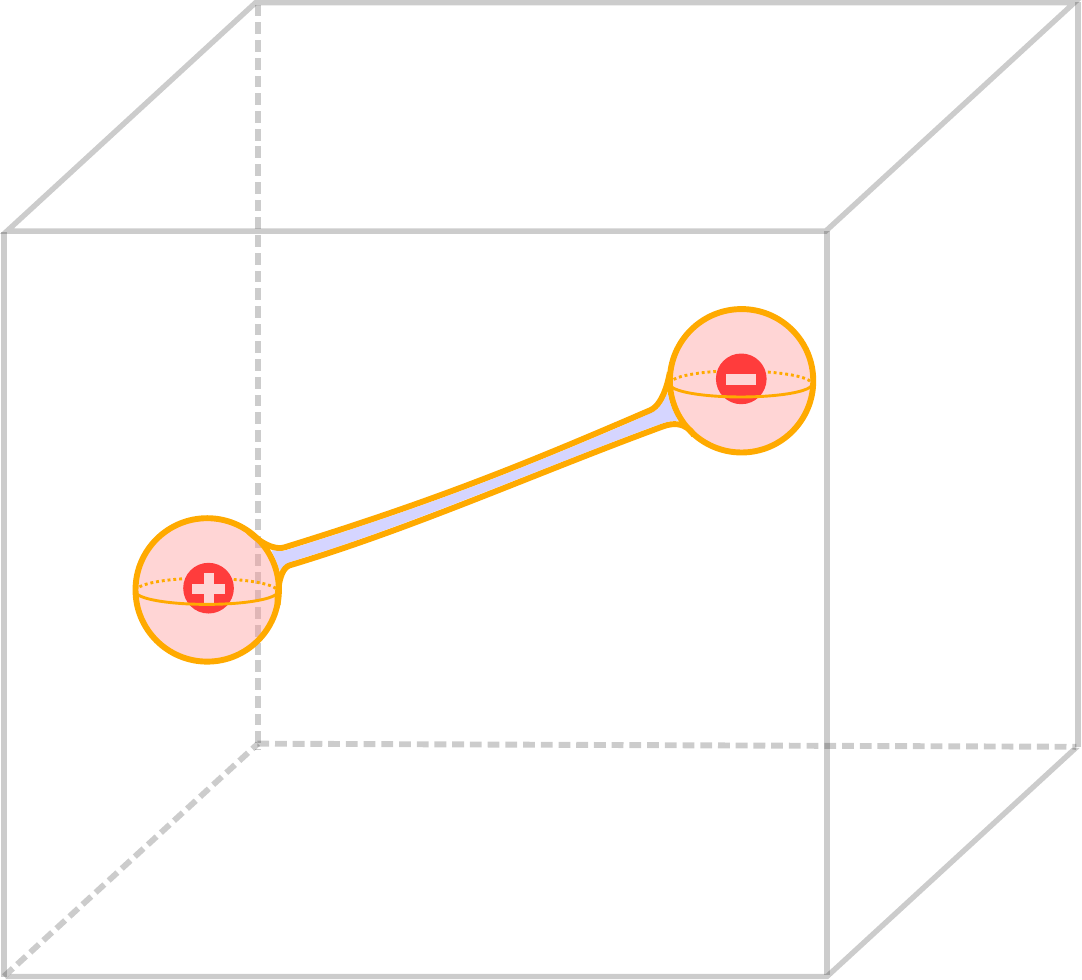}};
        \end{tikzpicture}
    \end{minipage}
\end{center}
\caption{(Left)
A schematic representation of the punctured manifold $M''$ with infinitesimally small $B_i$. In addition to $M'$, described in Fig.\,\ref{fig:thooft_loop_cut_off}, we indicate $B_i$ as blue regions. 
        (Right) A schematic picture when we take a time slice of the punctured manifold $M''$. The time slice is shown in the left figure as a gray slanted rectangle. In addition to the monopole and anti-monopole punctures as in Fig.\,\ref{fig:thooft_loop_cut_off}, $B_i$, which is the three-dimensional cylinder-like region connecting the monopole and anti-monopole punctures, is shown as a blue region. Note that, in this figure, $B_i$ looks like a straight cylinder, but just as Dirac strings, it can be any shape conguent to a cylinder, as long as it connects the monopole and anti-monopole punctures and infinitisimally thin.
        }
\label{fig:dirac_surface_cut_off}
\end{figure}

At this point, we can take whatever surface $S_i$ as long as it does not intersect the past and future hypersurfaces, but for the later convenience, we take $B_i$ to be as small as possible. To achieve this, we take $B_i$ to be very thin regions. See Fig.\,\ref{fig:dirac_surface_cut_off} for a schematic representation of the punctured manifold $M''$. The physical interpretation of $B_i$ is clear from the right figure of Fig.\,\ref{fig:dirac_surface_cut_off}: it encloses the Dirac string between the monopole and anti-monopole punctures on the time slice. In the limit of vanishing volume of $B_i$, the radius of the region surrounding the Dirac string is made to be infinitesimally small and we remove only the infinitesimal neighborhood of the Dirac string from the time slice. The rest of the spacetime manifold, $M''$, is the original spacetime manifold $M$ with the Dirac strings removed. It is obvious that the vector potential is well-defined on such a manifold.

Let us now consider the change of the helicity from the past hypersurface $\mathcal{S}_\text{past}$ to the future hypersurface $\mathcal{S}_\text{future}$, on which the helicity is defined as Eq.\,\eqref{eq:helicity_def_no_monopole_diff_form}.
On $M''$, the vector potential is well-defined and the field strength can be written as $F = dA$. Then, integrating $F \wedge F = d\qty(A \wedge F)$ over $M''$ yields
\begin{align}
    \int_{M''} F \wedge F = \mathcal{H}_\text{future} - \mathcal{H}_\text{past} + \sum_i \int_{S_i} A \wedge F,
    \label{eq:f_wedge_f_boundary_integral}
\end{align}
where we have used the Stokes theorem to rewrite the integral of $F \wedge F$ as the surface integral of $A \wedge F$.
The first two terms, $\mathcal{H}_\text{future}$ and $\mathcal{H}_\text{past}$ are the helicities at the infinite future and past, respectively, and comes from integrating $A \wedge F$ over the hypersurfaces $\mathcal{S}_\text{future}$ and $\mathcal{S}_\text{past}$.
The last term, $\sum_i \int_{S_i} A \wedge F$, is the contribution from $S_i=-\partial B_i$. 

With the infinitesimally small $B_i$, we may simplify the formula further. First, since by definition, the region $B_i$ does not include any singular regions, the integral of $F \wedge F$ over $B_i$ is vanishing. Therefore, we can rewrite the integral over $M''$ as
\begin{align}
    \int_{M''} F \wedge F = \int_{M'} F \wedge F.
\end{align}

Next, consider the integral
\begin{align}
    I_i = \int_{S_i} A \wedge F.
\end{align}
With the infinitesimally small $B_i$, we may separate the surface $S_i$ into two parts: the first part is (almost) the boundaries of the monopole and anti-monopole punctures. On a time slice, this corresponds to the spherical surfaces surrounding the monopole and anti-monopole punctures; see the right figure of Fig.\,\ref{fig:dirac_surface_cut_off}. The second part is the cylindrical surface connecting the monopole and anti-monopole punctures; in the right figure of Fig.\,\ref{fig:dirac_surface_cut_off}, it corresponds to the surface of the cylinder-like region connecting the monopole and anti-monopole punctures.
With the cylinder radius taken to zero, the integration regions of the first part can be regarded as $S^2$ surrounding the 't Hooft loop, and the second part can be regarded as $S^1$ surrounding the worldsheet of the Dirac string connecting the monopole and anti-monopole punctures.

According to this, let us separate the integral $I_i$ into two contributions:
one from the first part ($I_i^{\text{sphere}}$) and one from the second part ($I_i^{\text{cylinder}}$).
When we integrate $F$ over the sphere surrounding a monopole, we obtain
\begin{align}
    \int_{S^2} F = -\frac{2\pi}{e}.
\end{align}
Note that the minus sign for a monopole (not an anti-monopole) comes from our definition of the field strength tensor; see Eqs.\,\eqref{eq:field_strength_tensor} and \eqref{eq:electromagnetic_field} in App.\,\ref{app:notations}.
Then, when we integrate $A$ over the circle surrounding the Dirac string, we obtain
\begin{align}
    \int_{S^1} A = +\frac{2\pi}{e}.
    \label{eq:A_integral_dirac_string}
\end{align}
Here we take $S^1$ to be the clockwise direction when we see from the monopole side.
The integrals over other directions are all vanishing, and we obtain
\begin{align}
    I_i^{\text{sphere}} &= \int_{-\partial L_i} A \wedge F = - \frac{2\pi}{e}\int_{\partial \Sigma_i} A \\ 
    I_i^{\text{cylinder}} &= \int_{-S^1 \times \Sigma_i} A \wedge F = - \frac{2\pi}{e} \int_{\Sigma_i} F.
\end{align}
Here $L_i\subset B_i$ is topologically $S^1 \times B^3$ where $B^3$ is a three-dimensional ball containing the monopole on a time slice, and $S^1=\partial \Sigma_i$ is the $i$-th 't Hooft loop which is the boundary of the Dirac string worldsheet $\Sigma_i$. Then, we use $-\partial (S^1 \times B^3) = S^1\times \partial B^3 = \partial \Sigma_i\times S^2$ to compute the integral $I_i^{\text{sphere}}$\,\footnote{Note that, in general, we have $\partial(M\times N)= \partial M\times N \cup (-1)^{\text{dim}M}M\times \partial N$.}.
Similarly, the integration region appearing in $I_i^{\text{cylinder}}$ is $-\partial D^2\times \Sigma_i=-S^1 \times \Sigma_i$, where $D^2$ is a two-dimensional disk whose boundary is the circle $S^1$ surrounding the Dirac string on a time slice. The direction of $S^1$ is taken same as Eq.\,\eqref{eq:A_integral_dirac_string}.
Since there are no other 't Hooft loops inside the cylinder, we can write
\begin{align}
    I_i^{\text{sphere}} = - \frac{2\pi}{e} \int_{\Sigma_i} F = I_i^{\text{cylinder}}.
\end{align}
Therefore,
\begin{align}
    I_i = I_i^{\text{sphere}} + I_i^{\text{cylinder}} = - \frac{4\pi}{e} \int_{\Sigma_i} F.
\end{align}

This leads to the final expression for the helicity change:
\begin{align}
    \mathcal{H}_\text{future} = \mathcal{H}_\text{past} + \int_{M'} F \wedge F + \frac{4\pi}{e} \sum_i \int_{\Sigma_i} F. \label{eq:helicity_change}
\end{align}
The change of the helicity has additional contributions from the monopoles, compared to the helicity change formula without monopoles, Eq.\,\eqref{eq:helicity_change_no_monopole}.
Note that if $M$ has additional boundaries, we need to add the contributions from the boundaries to the right-hand side of Eq.\,\eqref{eq:helicity_change}.

Now, we can discuss the case where dyons, which are magnetic monopoles with electric charges, exist in the spacetime manifold $M$. We can follow the same procedure as above; if we focus on the regions except the Dirac strings, the vector potential $A_\mu$ is well-defined and the field strength $F_{\mu\nu}$ can be written as $F = dA$. The only difference is that, as the dyon has both electric and magnetic charges, the integration,
\begin{align}
    \int_{M'} F \wedge F,
\end{align}
diverges and sensitive to the choice of the UV cutoff scale, the radius of the ball $B^3$ surrounding the dyon worldline.
However, the helicity change formula is still well-defined and free from the divergence, as the UV divergence cancels out in the helicity change formula between the second and third terms in Eq.\,\eqref{eq:helicity_change}.

To see the cancellation explicitly, 
we consider the case where a dyon with magnetic charge $2\pi/e$ and electric charge $e$ is present in the spacetime manifold $M$. For simplicity, we take the topological angle, $\theta$, to be zero. Since it carries the elementary magnetic charge, the helicity change formula is the same as Eq.\,\eqref{eq:helicity_change} with the monopole contribution replaced by the dyon contribution. 
Let $\epsilon$ be the radius of the ball $B^3$ surrounding the dyon worldline and $L$ be the length of the dyon worldline. 
$\epsilon$ is the UV cutoff scale, which is taken to be much smaller than any length scale of the interest.
The divergent part of the integral $\int_{M'} F \wedge F$ is evaluated as follows:
\begin{align}
    \int_{M'} F \wedge F &= -2 L \int_{r>\epsilon} r^2 dr d\Omega \qty(\frac{e}{4\pi}) \qty(\frac{2\pi}{4\pi e}) \qty(\frac{1}{r^2})^2 + \mathcal{O}(\epsilon^0) \nonumber \nonumber \nonumber \\ 
    &= -\frac{L}{\epsilon} + \mathcal{O}(\epsilon^0)
    \label{eq:fft_divergent}
\end{align}
On the other hand, the contribution from the Dirac string worldsheet is evaluated as
\begin{align}
    \frac{4\pi}{e} \int_{\Sigma} F &= \frac{4\pi}{e} L \int_{r>\epsilon} \frac{e}{4\pi r^2} dr + \mathcal{O}(\epsilon^0) \nonumber \\
    &= \frac{L}{\epsilon} + \mathcal{O}(\epsilon^0).
\end{align}
In Eq.\,\eqref{eq:helicity_change}, the singular contributions from the second and third terms indeed cancel. In other words, the change of the helicity is a physical and well-defined quantity and free from the UV divergence. 
Furthermore, it should be independent of the regularization scheme such as the a sharp cutoff or a smooth smearing of the divergence at $r=0$.
This behavior is similar to the case of the chiral charge change of fermions in the presence of dyons, where the divergences also cancel between $\int F \wedge F$ and the contribution from the fermion number on the surfaces surrounding the dyon worldlines, as discussed in App.\,\ref{app:dyon_fermion_number}. 

It is important to note that the choice of $\Sigma_i$ is not unique. Indeed, discontinuous changes can occur if a Dirac string worldsheet crosses a monopole worldline. Nevertheless, the sum $\sum_i \int_{\Sigma_i} F$ remains invariant. To see the invariance, it is convenient to use another formalism to derive the helicity change formula, Eq.\,\eqref{eq:helicity_change}, which is valid even when the Dirac string worldsheets intersect each other. That is, we include the Dirac string worldsheets from the beginning\,\cite{Dirac:1948um}.
In this formalism, we may even generalize our result to generic $p$-form gauge theories with generalized helicity:
\begin{align}
    \mathcal{H} = \int A^{p_1} \wedge F^{p_2 + 1} \wedge \cdots \wedge F^{p_n + 1}, \label{eq:generalized_helicity}
\end{align}
where $A^{p_i}$ is the $p_i$-form vector potential and and $F^{p_j + 1} \equiv d A^{p_j}$, and $p_1 + p_2 + \cdots + p_n \leq D - n$, where $D$ is the spacetime dimension.

To begin with, let us first write the equations of motion of the $p$-form gauge theory in the presence of magnetic and electric branes in $D$ spacetime dimensions\,\cite{Fukuda:2020imw}:
\begin{align}
    \label{eq:elmg_eom}
    (-1)^{p+1} d \star F = e \delta_{M_e}, \quad
    d F = (-1)^{p(D - p)}g_m \delta_{M_m}, 
\end{align}
where $F$ is the field strength of the $p$-form gauge field, $e$ and $g_m$ are the electric and magnetic coupling constants, respectively, and $\delta_{M_e}$ and $\delta_{M_m}$ are the delta functions supported on the electric and magnetic branes.
The electric and magnetic coupling constants satisfy the Dirac quantization condition, $eg_m = 2\pi n$, where $n$ is an integer.
The delta function $\delta_M$ localized on the $n$-dimensional submanifold $M$ is defined as follows. First, we define the local coordinate system around $M$ as $(y^0, \ldots, y^{D-1})$ such that the worldvolume of $M$ is given by $y^n = y^{n+1} = \cdots = y^{D-1} = 0$. 
Then, the delta function is defined as
\begin{align}
    \delta_M = \delta(y^n) dy^n \wedge \cdots \wedge \delta(y^{D-1}) dy^{D-1}.
\end{align}
The integration of $n$-form $\omega$ over $M$ is now written as $\int_M \omega = \int \omega \wedge \delta_M$.

Reference \cite{Dirac:1948um} showed that (for $D=4$ and $p=1$) the second equation in Eq.\,\eqref{eq:elmg_eom} is solved by introducing the Dirac string worldsheet $\Sigma$, where $\partial\Sigma=M_m$, and the globally-defined vector potential $A_\mu$ as\,\cite{Hull:2024uwz}
\begin{align}
    F = dA - (-1)^{D(p + 1)} g_m \delta_\Sigma,
\end{align}
where $\delta_\Sigma$ is defined as $\int_{\Sigma} \omega = \int \omega \wedge \delta_\Sigma$ for any $D - p - 1$-form $\omega$.
This can be easily shown by noting that $d \delta_\Sigma = (-1)^{D - p - 1} \delta_{\partial \Sigma} = (-1)^{D - p - 1} \delta_{M_m}$, since for arbitrary $D - p - 2$-form $\omega$, $\int \omega \delta_{\partial \Sigma} = \int d\omega \delta_\Sigma$.
Indeed, $F$ satisfies the magnetic equation of motion, $d F = -g_m \delta_{M_m}$ for $D=4$ and $p=1$. 

The position of the Dirac string worldsheet $\Sigma$ is arbitrary, but it should satisfy the ``Dirac's veto'' condition\,\cite{Dirac:1948um,Brandt:1976hk,Hull:2024uwz}, which states that the Dirac string worldsheet $\Sigma$ should not intersect with the electric branes $M_e$. This is because the new vector potential $A_\mu$ is singular on $\Sigma$ and the equation of motion of the electric brane may be ill-defined if $\Sigma$ intersects with $M_e$. In our discussion, we do not consider the dynamics of the electric branes and may assume they are external objects, and we ignore this issue.

Now, we may re-derive the helicity change formula in Eq.\,\eqref{eq:helicity_change}.
On the future and past hypersurfaces, we may assume that no monopole worldlines or Dirac string worldsheets intersect the hypersurfaces. Then, the helicity on a hypersurface $\Sigma$ is given by $\int_\Sigma A \wedge dA$ and the change in helicity is given by
\begin{align}
    \mathcal{H}_\text{future} - \mathcal{H}_\text{past} &= \int_{M} dA \wedge dA
\end{align}
where $M$ is the four-dimensional spacetime manifold. Note that on the Dirac string worldsheet $\Sigma$, the vector potential $A_\mu$ is singular, as $\delta_\Sigma$ is singular but $F$ is well-defined. For the time being, we may assume that the delta-function singularity is regularized by properly smearing it\,\cite{Dirac:1948um,Hull:2024uwz} and the integral is well-defined. 
Then, we can rewrite the integral using $F$ and $\delta_\Sigma \equiv \sum_i \delta_{\Sigma_i}$ as
\begin{align}
    \mathcal{H}_\text{future} - \mathcal{H}_\text{past} &= \int_{M} \qty(F + g_m \delta_\Sigma) \wedge \qty(F + g_m \delta_\Sigma) \nonumber \\
    &= \int_{M} F \wedge F + 2g_m \sum_i \int_{\Sigma_i} F + 2g_m^2 N_{\#},
\end{align}
where $N_{\#}\equiv\sum_{i<j}\int_M\delta_{\Sigma_i}\wedge\delta_{\Sigma_j}$ is the signed intersection number of the Dirac string worldsheets $\Sigma_i$ and $\Sigma_j$ for $i \neq j$\,\footnote{The computation of the self-intersection number corresponding to $i=j$ may require the regularization. Here we take $\delta_{\Sigma_i}\wedge \delta_{\Sigma_i}=0$ which is consistent with the earlier derivation of \eqref{eq:helicity_change}.}. Recalling the Dirac quantization condition $eg_m = 2\pi$, we have
\begin{align}
    \label{eq:helicity_change_generic}
    \mathcal{H}_\text{future} = \mathcal{H}_\text{past} + \int_{M} F \wedge F + \frac{4\pi}{e} \sum_i \int_{\Sigma_i} F + 2\qty(\frac{2\pi}{e})^2 N_{\#}.
\end{align}
This indeed reproduces the helicity change formula in Eq.\,\eqref{eq:helicity_change} when the Dirac string worldsheet $\Sigma_i$ is chosen so that the intersection number $N_{\#}$ is vanishing, i.e., the worldsheets do not intersect each other.
The final formula does not depend on the regularization of the delta-function singularity.
This formalism can be easily extended to generic $D$ spacetime dimensions and $p$-form gauge theories\,\cite{Hull:2024uwz} and we can calculate the change in the generalized magnetic helicity, Eq.\,\eqref{eq:generalized_helicity}.

With the addition of the intersection number term, we can now easily show that the helicity change, or the last two terms in the right-hand side of Eq.\,\eqref{eq:helicity_change_generic}, is invariant even if we choose a different choise of the Dirac string worldsheet $\Sigma_i$. Let us define $\mathcal{I}$ as
\begin{align}
    \mathcal{I} \equiv 2g_m \sum_i \int_{\Sigma_i} F + 2g_m^2 N_{\#}
\end{align}
and suppose we change one of the Dirac string worldsheet $\Sigma_i$ to $\Sigma_i'$. If we continuously deform $\Sigma_i$, i.e., it does not cross over any monopole worldlines, the intersection number $N_{\#}$ does not change and
\begin{align}
    \Delta \mathcal{I} = 2g_m \int_{\Sigma_i} F - 2g_m \int_{\Sigma_i'} F = 2g_m \oint_{\Sigma_i \cup (-\Sigma_i')} F = 0,
\end{align}
as the closed surface $\Sigma_i \cup (-\Sigma_i')$ 
does not enclose any monopole worldlines. On the other hand, if we discontinuously change the Dirac string worldsheet $\Sigma_i$ to $\Sigma_i'$, i.e., the Dirac string worldsheet crosses over a monopole worldline,
in such a way that the common surface $\Sigma_i \cup (-\Sigma_i')$ positively encloses a monopole worldline,
the intersection number $N_{\#}$ changes by $1$, and we have
\begin{align}
    \Delta \mathcal{I} = 2g_m \int_{\Sigma_i} F - 2g_m \int_{\Sigma_i'} F + 2g_m^2 \Delta N_{\#} = 2g_m \oint_{\Sigma_i \cup (-\Sigma_i')} F + 2g_m^2 = 0,
\end{align}
as $\oint_{\Sigma_i \cup (-\Sigma_i')} F = -g_m$.

Let us now discuss the physical meaning of the helicity change formula in Eq.\,\eqref{eq:helicity_change}. The first two terms, $\mathcal{H}_\text{past} + \int_{M'} F \wedge F$, represent the change in helicity arising from the evolution of the electromagnetic field in the absence of monopoles. In the context of chiral anomalies, this combination corresponds to the change in chiral charge.
The final term, $\frac{2\pi}{e} \sum_i \int_{\Sigma_i} F$, captures the contribution to the helicity change due to the presence of monopoles. Physically, this term quantifies the amount of helicity that is effectively ``removed'' or ``torn apart'' by the monopoles as they traverse the system.
In other words, the difference between the change in helicity and $\int_{M'} F \wedge F$ is precisely accounted for by the monopole contribution. In the following, we illustrate this interpretation with explicit examples.

\subsection{Example: Helicity change with a single monopole}
\label{sec:helicity_change_example_A}

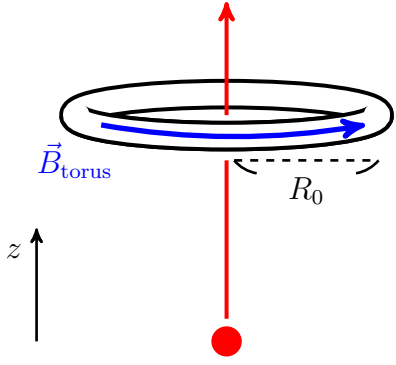
\begin{figure}[t]
\begin{center}
    \begin{tikzpicture}
    \pgfsetlinewidth{1.5pt}
 \pic{torus={2cm}{1.5mm}{82}};
    \draw[blue,line width=2pt,->,>=stealth'] (-1.65cm, -0.13cm) arc [radius=10cm, start angle=260, end angle=280];
    \node[blue] at (-2cm, -0.6cm) {\large $\vec{B}_\text{torus}$};
    \fill[red] (0, -3cm) circle (0.2cm);
    \draw[red, line width=1.5pt] (0, -2.7cm) -- (0, -0.6cm);
    \draw[red, line width=1.5pt, ->,>=stealth'] (0, 0) -- (0, 1.5cm);
    \draw[line width=1pt, ->,>=stealth'] (-2.5cm, -3cm) -- (-2.5cm, -1.5cm);
    \node at (-2.8cm, -1.8cm) {\large $z$};
    \draw[dashed, line width=1pt] (0.1cm, -0.6cm) -- (2.0cm, -0.6cm);
    \draw[line width=1pt] (0.1cm, -0.6cm) arc [radius=0.5cm, start angle=220, end angle=260];
    \draw[line width=1pt] (2.0cm, -0.6cm) arc [radius=0.5cm, start angle=320, end angle=280];
    \node at (1.05cm, -1cm) {\large $R_0$};

\end{tikzpicture}
\end{center}
\caption{A toroidal magnetic flux with a monopole moving along the $z$-axis. The toroidal magnetic flux is centered at the origin $O$, with radius $R_0$. The monopole is represented by the red dot at the bottom, moving upwards along the $z$-axis.}
\label{fig:helicity_change_example_A}
\end{figure}

Consider a configuration where a toroidal (donut-shaped) magnetic field carries a total magnetic flux $\Phi_0$. In this initial state, the helicity is zero. Now, suppose a magnetic monopole--anti-monopole pair is created somewhere in spacetime. The monopole is transported through the center of the toroidal flux and then annihilates with the anti-monopole. The helicity remains zero both in the infinite past and future. However, during this process, the integral $\int_{M'} F \wedge F$ becomes nonzero:
\begin{align}
    \label{eq:fwedgef_exA}
    \int_{M'} F \wedge F = \frac{4\pi}{e} \Phi_0.
\end{align}
Let us check this explicitly. First, for simplicity, we put the toroidal flux in the $xy$-plane at $z=0$, and move a monopole along the $z$-axis with a non-relativistic constant velocity $v$; the monopole coordinates are $(0, 0, vt)$.
We assume the radius of the toroidal flux is $R_0$, which is much larger than the small radius of the toroidal flux. See Fig.\,\ref{fig:helicity_change_example_A} for the configuration.

The magnetic field in the comoving frame of the monopole is given by
\begin{align}
    \vec{B}' = \frac{g_m}{4\pi} \frac{\vec{r}}{r^3}
\end{align}
where $g_m$ is the unit magnetic charge, $\vec{r}$ is the position vector in the comoving frame.
In the lab frame, $\vec{r} = (x, y, z - vt) = \rho \,\vec{e}_\rho + (z - vt) \vec{e}_z$ in the cylindrical coordinates, $(\rho, \theta, z)$.
The electric field in the lab frame arises from the Lorentz transformation of the magnetic field with the boost velocity $\beta = -v$ (see Eq.~\eqref{eq:Lorentz_transformation_E}):
\begin{align}
    \vec{E} &= -\vec{v} \times \vec{B}' \nonumber \\ 
    &= -v \vec{e}_z \times \frac{g_m}{4\pi} \frac{\rho \vec{e}_\rho + (z - vt) \vec{e}_z}{(\rho^2 + (z - vt)^2)^{3/2}}  \nonumber \\
    &= - \frac{g_m}{4\pi} \frac{v \rho}{(\rho^2 + (z - vt)^2)^{3/2}} \vec{e}_\theta.
\end{align}
Now, except for the electric field from the moving monopole, there is no electric field in the lab frame. Therefore, $E\cdot B$ arises only from the toroidal flux and the electric field from the monopole:
\begin{align}
    \int F \wedge F &= -2\int dt d^3x \vec{E} \cdot \vec{B} \nonumber \\
    &= - 2\int dt d^3x \qty(-\frac{g_m}{4\pi}) \frac{v \rho}{(\rho^2 + (z - vt)^2)^{3/2}} B_\text{torus}.
\end{align}
Here, $B_\text{torus}$ is the magnetic field of the toroidal flux, $\vec{B}_\text{torus}=B_\text{torus}\vec{e}_\theta$. As we assume the torus is thin, we can approximate the magnetic field as $B_\text{torus} = \delta(\rho - R_0) \delta(z) \Phi_0$. Then,
\begin{align}
    \int F \wedge F &= \frac{g_m}{2\pi} \Phi_0 v \cdot 2\pi R_0^2 \int_{-\infty}^{\infty} \frac{dt}{(R_0^2 + v^2 t^2)^{3/2}} \nonumber \\
    &= 2 g_m \Phi_0.
\end{align}
Since the unit magnetic charge is $g_m = \frac{2\pi}{e}$ from the Dirac quantization condition, we have Eq.\,\eqref{eq:fwedgef_exA}.

Meanwhile, the integral over the Dirac string worldsheet, which measures the magnetic flux passing through it, is also nonzero:
\begin{align}
    \int_{\Sigma} F = -\Phi_0.
\end{align}
Again, one must be careful about the definition of the field strength tensor, Eq.\,\eqref{eq:field_strength_tensor}, for the sign.
Substituting into the helicity change formula~\eqref{eq:helicity_change}, we obtain
\begin{align}
    \mathcal{H}_\text{future} = \mathcal{H}_\text{past} + \int_{M'} F \wedge F + \frac{4\pi}{e} \int_{\Sigma} F = 0 + \frac{4\pi}{e} \Phi_0 - \frac{4\pi}{e} \Phi_0 = 0,
\end{align}
which confirms that the net helicity remains unchanged, as expected.

\subsection{Example: Helicity change in the Higgs phase}
\label{sec:helicity_change_example_B}
\begin{figure}[t]
\begin{center}
    \includegraphics[scale=0.5]{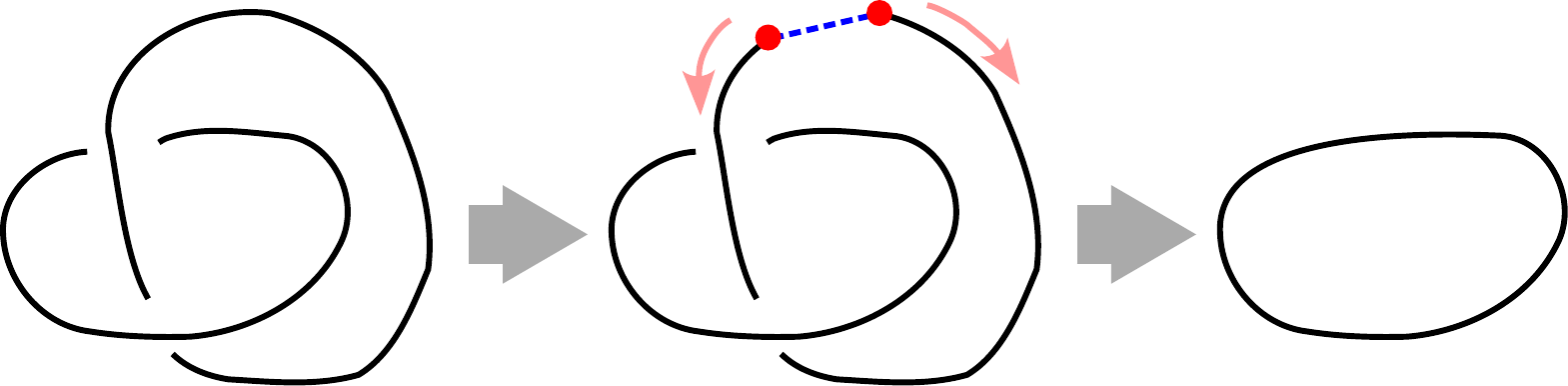}
\end{center}
\caption{A schamatic illustration of the helicity change in the Higgs phase. At the initial state, two linked vortex loops are present, and the helicity is nonzero, as shown in the leftmost figure. At some point in time, a monopole--anti-monopole pair is nucleated, as shown in the middle figure. The monopole and anti-monopole are represented by the red dots and a possible choise of the Dirac string between them is shown as the blue dashed line.
 They move along the vortex loop as shown by the pink arrows in the middle figure. They eventually annihilate, leaving only one vortex loop in the final state, as shown in the rightmost figure. The helicity is zero in the final state.}
\label{fig:helicity_change_example_B}
\end{figure}

Next, consider the theory in the Higgs phase. To realize a nonzero helicity, arrange two Abrikosov--Nielsen--Olesen vortex loops so that they are linked in space. In this configuration, the initial helicity is
\begin{align}
    \mathcal{H}_\text{past} = 2 \qty(\frac{2\pi}{e})^2.
\end{align}
Suppose now that one vortex is ``torn apart'' by nucleating a monopole--anti-monopole pair, which then annihilate, leaving only a single vortex loop in the final state. The final helicity is thus zero. The schematic illustration of this process is shown in Fig.\,\ref{fig:helicity_change_example_B}.

Let us evaluate the relevant integrals. $\int_{M'} F \wedge F$ vanishes, as $E=0$ in the Higgs phase.
The Dirac string worldsheet integral measures the flux of the remaining vortex:
\begin{align}
    \int_{\Sigma} F = -\frac{2\pi}{e}.
\end{align}
Thus, the helicity change is
\begin{align}
    \mathcal{H}_\text{future} &= \mathcal{H}_\text{past} + \int_{M'} F \wedge F + \frac{4\pi}{e} \int_{\Sigma} F \\
    &= 2 \qty(\frac{2\pi}{e})^2 - 0 - \frac{4\pi}{e} \cdot \frac{2\pi}{e} = 0.
\end{align}
This confirms that the helicity vanishes in the final state, consistent with the physical expectation.


\section{Helicity change in the Georgi--Glashow model}
\label{sec:non_abelian_adjoint_Higgs}
In the previous section, we discussed the helicity change in the effective Abelian gauge theory. We have found that the change of the helicity is well-defined, but it is modified from the conventional helicity change formula without monopoles, Eq.\,\eqref{eq:helicity_change_no_monopole}, by the contribution from the magnetic fields on the Dirac string worldsheet.

Then, it is quite natural to ask what happens if we embed the IR effective Abelian gauge theory in a non-Abelian gauge theory. In this section, we will discuss the helicity change in the Georgi--Glashow model, which is a non-Abelian gauge theory with an adjoint Higgs field. 
In the vacuum, the adjoint Higgs field $\Phi^a$ acquires a vacuum expectation value (VEV) $\langle \Phi^a \rangle = v \delta^{a3}$. This breaks the gauge symmetry $\text{SU}(2)$ down to $\text{U}(1)$, and the gauge field $A_\mu^3$ out of the original $\text{SU}(2)$ gauge field $A^a_\mu$ becomes the electromagnetic field. The theory has a topological soliton solution, the 't Hooft--Polyakov monopole\,\cite{tHooft:1974kcl,Polyakov:1974ek}, which enjoys a unit magnetic charge in the low-energy $\text{U}(1)$ gauge theory.

The key lesson from the previous section is that, in the presence of magnetic monopoles, the gauge potential $A_\mu$ is not globally defined, so the naive helicity change formula, Eq.\,\eqref{eq:helicity_change_no_monopole}, cannot be applied as is. One might suspect that the Georgi--Glashow model evades this issue: in Minkowski spacetime with a trivial principal $\text{SU}(2)$ bundle --- i.e., in the absence of instantons\,\footnote{Instantons are defined in Euclidean spacetime.} --- the non-Abelian potential $A^a_\mu$ can be chosen globally even when the adjoint Higgs $\Phi^a$ has a nontrivial profile supporting monopoles. Nevertheless, essentially the same modification as in the Abelian case, Eq.\,\eqref{eq:helicity_change}, still arises: the obstruction reappears after descending to the low-energy $\text{U}(1)$ theory, because its effective gauge potential need not be globally well defined even though the UV $\text{SU}(2)$ potential is.

Let us first write down the IR effective Abelian vector potential $\mathcal{A}_\mu$ in the Georgi--Glashow model with non-vanising Higgs VEV. In the trivial vacuum, $\langle \Phi^a \rangle = v \delta^{a3}$, the vector potential is simply given by
\begin{align*}
    \mathcal{A}_\mu = A_\mu^3.
\end{align*}
Therefore, one may wonder that even with generic non-constant Higgs field, the vector potential $\mathcal{A}_\mu$ can be defined as the projection of the $\text{SU}(2)$ gauge potential $A_\mu^a$ onto the local direction of the Higgs field, i.e., 
\begin{align}
    \label{eq:wrong_abelian_vector_potential}
    \mathcal{A}_\mu \stackrel{?}{=} A_\mu^a n^a,
\end{align}
where $n^a$ is the normalized Higgs field,
\begin{align}
    n^a = \frac{\Phi^a}{\sqrt{|\Phi|^2}}.
\end{align}
However, this does not work in general; the obvious pathological problem is that the expression, Eq.\,\eqref{eq:wrong_abelian_vector_potential}, is not gauge invariant under the $\text{SU}(2)$ gauge transformation even up to the gauge transformation in the IR $\text{U}(1)$ gauge theory; under the gauge transformation (see Eq.~\eqref{eq:non-abelian_gauge_transformation} for the transformation law of the gauge field), 
\begin{align}
    A_\mu^a n^a = 2 \text{Tr} \qty(A_\mu n) \to A_\mu^a n^a + \frac{2i}{g} \text{Tr} \qty(\partial_\mu U^\dagger U n),
\end{align}
where $A_\mu \equiv A_\mu^a T^a$, $n \equiv n^a T^a$, $T^a$ is the generator of $\text{SU}(2)$ in the fundamental representation, $U$ is the gauge transformation in $\text{SU}(2)$ in the fundamental representation, and $g$ is the gauge coupling constant.
In other words, \emph{there is no {\it a priori} well-defined vector potential in the low-energy $\text{U}(1)$ gauge theory}.

Then, what should be the well-defined quantity in the low-energy $\text{U}(1)$ gauge theory in terms of the UV $\text{SU}(2)$ gauge theory?
It should be a gauge invariant quantity not only in the UV $\text{SU}(2)$ gauge theory but also in the IR $\text{U}(1)$ gauge theory. The simplest candidate is the electromagnetic field strength $\mathcal{F}_{\mu\nu}$; how can we define it in terms of the UV $\text{SU}(2)$ gauge theory?
Reference \cite{tHooft:1974kcl} shows that the following tensor is well-defined and can be reduced to the electromagnetic field strength in the vacuum, $\langle \Phi^a \rangle = \text{const.}$:
\begin{align}
    \label{eq:thooft_tensor}
    &\mathcal{F} = \frac{1}{2}\mathcal{F}_{\mu\nu}dx^\mu\wedge dx^\nu,
    &&\mathcal{F}_{\mu\nu} = F_{\mu\nu}^a n^a - \frac{1}{g} \epsilon^{abc} n^a D_\mu n^b D_\nu n^c,
\end{align}
where $F_{\mu\nu}^a$ is the field strength of the $\text{SU}(2)$ gauge theory (defined in Eq.~\eqref{eq:non-abelian_field_strength}) and $D_\mu n^a = \partial_\mu n^a + g \epsilon^{abc} A_\mu^b n^c$ is the covariant derivative of $n$ (see also Ref.\,\cite{Weinberg:1996kr}). This tensor is sometimes called the 't Hooft tensor. Obviously, $\mathcal{F}_{\mu\nu}$ is gauge invariant under the $\text{SU}(2)$ gauge transformation. It is also closed and is reduced to the electromagnetic field strength in the vacuum, $\langle \Phi^a \rangle = v \delta^{a3}$. To see this, one may first note that $\mathcal{F}_{\mu\nu}$ can be expressed in the following way\,\cite{Weinberg:1996kr}:
\begin{align}
    \label{eq:thooft_tensor_reduced}
    \mathcal{F}_{\mu\nu} = \partial_\mu \qty(n^a A_\nu^a) - \partial_\nu \qty(n^a A_\mu^a) - \frac{1}{g} \epsilon^{abc} n^a \partial_\mu n^b \partial_\nu n^c.
\end{align}
From this expression, it is clear that $\mathcal{F}_{\mu\nu}$ reduces to the electromagnetic field strength in the vacuum. The Bianchi identity can be checked as follows:
\begin{align}
    \varepsilon^{\mu\nu\rho\sigma} \partial_\nu \mathcal{F}_{\rho\sigma} = -\frac{1}{g} \varepsilon^{\mu\nu\rho\sigma} \epsilon^{abc} \partial_\nu n^a \partial_\rho n^b \partial_\sigma n^c.
\end{align}
Since $n^a$ is a unit vector, $\partial_\mu n^a$ is orthogonal to $n^a$: $n^a \partial_\mu n^a = 0$. The right-hand side is zero because it antisymmetrizes three vectors in the two-dimensional subspace orthogonal to $\vec n$.

The Bianchi identity is equivalent to the statement that
$\mathcal{F}$ is a closed two-form. By the Poincar\'e lemma, it is locally exact, i.e., there exists a one-form $\mathcal{A}_\mu$ such that $\mathcal{F} = d\mathcal{A}$. This one-form $\mathcal{A}_\mu$ is the vector potential in the low-energy $\text{U}(1)$ gauge theory. In other words, the IR vector potential is \emph{derived {\it a posteriori} from the 't Hooft tensor $\mathcal{F}_{\mu\nu}$} in the low-energy $\text{U}(1)$ gauge theory.
Note that,
although the original $\text{SU}(2)$ gauge field $A_\mu^a$ is globally well-defined and the Higgs field $n^a$ is also well-defined unless on the monopole worldlines, the vector potential $\mathcal{A}_\mu$ in the low-energy $\text{U}(1)$ gauge theory may not be globally well-defined in general\,\footnote{
    One may still wonder if it is possible that the principal $\text{U}(1)$ bundle, which is the subbundle of the trivial principal $\text{SU}(2)$ bundle, is non-trivial. This is completely legitimate. An elementary example is the M\"obius strip: suppose a $\mathbf{R}^2$ is trivially fibered over the circle $S^1$. We may take a subspace of the fiber $\mathbf{R}^2$, a one-dimensional straight line passing through the origin. Using the coordinate of the base space $S^1$, $\theta \in [0, 2\pi)$, we take the direction of the line to be $\vec{e}_\theta = (\cos \theta / 2, \sin \theta / 2)$; the line is parametrized as $t \vec{e}_\theta$ for $t \in \mathbf{R}$. If we restrict $t \in [-1, 1]$, we obtain a M\"obius strip, which is a non-trivial fiber bundle over $S^1$. Thereby, we may obtain a non-trivial bundle from a trivial bundle by taking a subspace of the fiber. 
}. The condition for the vector potential $\mathcal{A}_\mu$ to be well-defined is, as we have discussed in Sec.\,\ref{sec:abelian_helicity_change}, that the Chern number of the low-energy $\text{U}(1)$ gauge theory is vanishing, i.e., the monopole does not exist at all. 
We are now ready to simplify the 't Hooft tensor and derive the relation between the IR helicity and UV quantities.
Before proceeding, let us comment on how the electromagnetic field strength $\mathcal{F}_{\mu\nu}$ is constructed from the UV $\text{SU}(2)$ gauge theory in the literature.
In some literatures
such as Refs.\,\cite{Vachaspati:1991nm,Grasso:1997nx,Grasso:2000wj,Kandus:2010nw,GarciaGarcia:2025uub}, they use the following definition:
\begin{align}
    \mathcal{F}'_{\mu\nu} = F_{\mu\nu}^a n^a - \frac{1}{g|\Phi|^3} \epsilon^{abc} \Phi^a D_\mu \Phi^b D_\nu \Phi^c.
\end{align}
This expression looks different from ours, Eq.\,\eqref{eq:thooft_tensor} at first glance, but is equivalent\,\footnote{We would like to thank 
Shuailiang Ge for pointing this out.} and both are closed\,\footnote{
Ref.\,\cite{GarciaGarcia:2025uub} suggests that the IR field strength should ``violate the standard $\text{U}(1)$ Bianchi identity even though $A^a_\mu$ does not exhibit singular behavior.'' However, this could be misleading. As discussed above, the apparent ``singular behavior'' of the IR vector potential arises from the need to patch local charts when applying the Poincar\'e lemma, which in turn reflects the fact that $n^a$ and thus $\mathcal{F}_{\mu\nu}$ are ill-defined where $\langle \Phi^a \rangle = 0$. In other words, even in the presence of monopoles (with $B \sim 1/r^2$), $\text{div} \vec{B} = 0$ holds everywhere except on the monopole worldlines, and the Bianchi identity is preserved away from these singularities.
}, as $D_\mu n^a = (D_\mu \Phi^a) / |\Phi| + \Phi^a \partial(1/|\Phi|)$ and therefore 
\begin{align*}
    \frac{1}{|\Phi|^3} \epsilon^{abc} \Phi^a D_\mu \Phi^b D_\nu \Phi^c = \epsilon^{abc} n^a D_\mu n^b D_\nu n^c.
\end{align*}

Let us now further simplify the 't Hooft tensor $\mathcal{F}_{\mu\nu}$. In matrix and differential form notation, Eq.~\eqref{eq:thooft_tensor_reduced} can be written as
\begin{align}
    \label{eq:thooft_tensor_matrix}
    \mathcal{F} 
    = 2\, d\, \mathrm{Tr}(n A) + \frac{2i}{g}\, \mathrm{Tr}(n\, d n \wedge d n),
\end{align}
where $A=A_\mu^a T^a dx^\mu$, and we have used
\begin{align}
    &\mathrm{Tr}\left(T^a T^b\right)=\frac{1}{2}\delta^{ab}, 
    &&\mathrm{Tr}\left(T^a T^b T^c\right)
    =\frac{i}{4}\epsilon^{abc}.
\end{align}
Since $\mathcal{F}$ is gauge-invariant, we are free to choose a gauge where the Higgs direction $n$ is constant in spacetime, i.e., $n \to U n U^\dagger = T^\Xi$, with $\Xi$ a fixed direction (e.g., $\Xi=3$) and $T^\Xi$ the corresponding generator. 
The gauge field transforms as in Eq.~\eqref{eq:non-abelian_gauge_transformation}. 
In this gauge, the 't Hooft tensor becomes
\begin{align}
    \label{eq:thooft_tensor_matrix_reduced}
    \mathcal{F} = 2\, d\, \mathrm{Tr}(n A) + \frac{2i}{g}\, d\, \mathrm{Tr}(T^\Xi U d U^\dagger).
\end{align}
By comparing Eqs.~\eqref{eq:thooft_tensor_matrix} and \eqref{eq:thooft_tensor_matrix_reduced}, we observe
\begin{align}
    \frac{i}{g}\, \mathrm{Tr}(n\, d n \wedge d n) = \frac{i}{g}\, d\, \mathrm{Tr}(T^\Xi U d U^\dagger) = \frac{i}{g}\mathrm{Tr}(T^\Xi \alpha \wedge \alpha),
\end{align}
where $\alpha \equiv i^{-1} U d U^\dagger \equiv T^a \alpha^a$ is the Maurer--Cartan one-form associated with the gauge transformation $U$.
From Eq.\,\eqref{eq:thooft_tensor_matrix_reduced}, the corresponding vector potential $\mathcal{A}$ (with $\mathcal{F} = d\mathcal{A}$) is
\begin{align}
    \mathcal{A} = n^a A^a - \frac{2}{g}\, \mathrm{Tr}(T^\Xi \alpha),
\end{align}
up to a $\text{U}(1)$ gauge transformation.
For notational simplicity, we introduce the one-form $\omega$ as
\begin{align}
    &\omega = \frac{1}{g} \mathrm{Tr}(T^\Xi \alpha),
    &&d\omega 
    = -\frac{i}{g}\mathrm{Tr}(T^\Xi \alpha \wedge \alpha)
    = - \frac{i}{g}\, \mathrm{Tr}(n\, d n \wedge d n) 
\end{align}
so that
\begin{align}
    \label{eq:A_F_georgi_glashow}
    &\mathcal{A} = n^a A^a - 2\omega,
    &&\mathcal{F} = d\,(n^a A^a) - 2d\omega.
\end{align}
It is important to note that this construction is only valid locally: in the presence of monopoles, the gauge transformation $U$ cannot be defined globally. Then, $\omega$ is not globally well-defined, although $d\omega$ is.
This obstruction is equivalent to the impossibility of globally ``combing the hedgehog'' configuration of the Higgs field in the 't Hooft--Polyakov monopole background~\cite{Shifman:2012zz}. In such cases, the singularity in the gauge transformation $U$ manifests as a Dirac string in the vector potential, reflecting the topological obstruction to defining $\mathcal{A}$ everywhere.

With this explicit form for the vector potential, we can express the IR helicity in terms of UV gauge-theoretic quantities, as anticipated in Ref.\,\cite{Hamada:2025cwu}. 
To discuss the helicity, we need to assume that no monopole is present in the spatial surface of interest, so that we may comb the Higgs field globally; the Maurer--Cartan one-form $\alpha$ is globally well-defined, and so is $\omega$.
The IR helicity on a spatial slice $\mathcal{S}$ is then
\begin{align}
    \mathcal{H}_\text{IR} = \int_{\mathcal{S}} \mathcal{A} \wedge \mathcal{F}.
\end{align}
Expanding this using the 't Hooft tensor, we find (by using Eq.~\eqref{eq:A_F_georgi_glashow} for $\mathcal{A}$ and Eq.~\eqref{eq:thooft_tensor} for $\mathcal{F}$)
\begin{align}
    \mathcal{A} \wedge \mathcal{F} &= n^a n^b A^a \wedge F^{b} - (n^a A^a) \wedge \frac{1}{2} \eta - 2\omega \wedge (n^a F^{a}) + \omega \wedge \eta,
\end{align}
where
\begin{align}
    \eta &\equiv \frac{1}{g} \epsilon^{abc} n^a D n^b \wedge D n^c 
    = 4 d\omega + 2 A^a \wedge d n^a + g \epsilon^{abc} A^a \wedge A^b n^c.
\end{align}
Using the identity
\begin{align}
    A^a \wedge A^b \wedge A^c = \epsilon^{abc} A^1 \wedge A^2 \wedge A^3 = \frac{1}{6} \epsilon^{abc} \epsilon^{def} A^d \wedge A^e \wedge A^f,
\end{align}
the helicity can be further rewritten as
\begin{align}
    \label{eq:helicity_ir_georgi_glashow}
    \mathcal{H}_\text{IR} = \int_{\mathcal{S}} \left[
        \mathcal{K} + 4\omega \wedge d\omega
        - (\delta^{ab} - n^a n^b) A^a \wedge F^b
        - \frac{1}{g} n^a A^a \wedge (\epsilon^{bcd} n^b D n^c \wedge d n^d)
    \right],
\end{align}
where $\mathcal{K}$ is the non-Abelian Chern--Simons three-form defined in Eq.\,\eqref{eq:non-abelian_cs}.

Although none of the individual terms on the right-hand side of Eq.\,\eqref{eq:helicity_ir_georgi_glashow} is gauge invariant (only the sum is), it is still useful to discuss their separate physical meanings.
The first term, $\mathcal{K}$, gives the non-Abelian Chern--Simons number,
\begin{align}
    N_\text{CS} = \frac{g^2}{16\pi^2} \int \mathcal{K}.
\end{align}
For the second term, note that
\begin{align}
    \Omega \equiv \frac{g}{2\pi} d\omega = \frac{1}{8\pi} \epsilon^{abc} n^a d n^b \wedge d n^c 
\end{align}
is the pullback of the normalized volume form on $S^2 \cong \text{SU}(2)/\text{U}(1)$ by the Higgs unit vector $n^a$. Hence,
\begin{align}
    H_\Phi \equiv \frac{g^2}{16 \pi^2} \int_{\mathcal{S}} 4 \omega \wedge d\omega,
    \label{eq:hopf_invariant}
\end{align}
if the spatial slice $\mathcal{S}$ is congruent to $S^3$, 
is the Hopf invariant\,(Sec.~9.4 in Ref.\,\cite{Nakahara:2003nw}), i.e., the winding number in $\pi_3(S^2) \cong \mathbf{Z}$. Explicitly,\footnote{In the computation, index $\Xi$ is not summed over while indices $a,b,c$ are summed over.}
\begin{align}
    \frac{16 \pi^2}{g^2} H_\Phi = \int 4\omega \wedge d\omega &= -\frac{4i}{g^2} \int \mathrm{Tr}(T^\Xi \alpha) \wedge \mathrm{Tr}(T^\Xi \alpha \wedge \alpha) \nonumber \\ 
    &= \frac{4}{g^2} \frac{1}{8} \int \alpha^\Xi \wedge \epsilon^{\Xi bc} \alpha^b \wedge \alpha^c \nonumber \\
    &= \frac{4}{g^2} \frac{1}{24} \int \epsilon^{abc} \alpha^a \wedge \alpha^b \wedge \alpha^c \nonumber \\
    &= -\frac{4i}{g^2} \frac{1}{6} \int \text{Tr} \bigl(\alpha \wedge \alpha \wedge \alpha \bigr) \nonumber \\
    &= \frac{16 \pi^2}{g^2} N_H,
\end{align}
with the Higgs winding number
\begin{align}
    N_H \equiv \frac{1}{24 \pi^2} \int \text{Tr}(U d U^\dagger \wedge U d U^\dagger \wedge U d U^\dagger).
\end{align}
Therefore, if the third and fourth terms in Eq.\,\eqref{eq:helicity_ir_georgi_glashow} vanished, we would obtain
\begin{align}
    \label{eq:helicity_ir_conjectured}
    \mathcal{H}_\text{IR} = \frac{16\pi^2}{g^2} \bigl(N_\text{CS} + N_H\bigr),
\end{align}
which is the relation conjectured in Ref.\,\cite{Hamada:2025cwu}.

We now interpret the remaining third and fourth terms.
When the Higgs direction is covariantly constant, $D_\mu n^a = 0$, the fourth term is zero. Moreover,
\begin{align}
    D_\mu n^a = 0 \;\Rightarrow\; [D_\mu, D_\nu] n^a = 0 \;\Leftrightarrow\; \epsilon^{abc} F^a_{\mu\nu} n^b = 0.
\end{align}
This states that $F^a_{\mu\nu}$ is parallel to $n^a$, i.e., $(\delta^{ab} - n^a n^b) F^b_{\mu\nu} = 0$, which means that the perpendicular (heavy) gauge modes have decoupled from the IR sector\,\footnote{The converse is not true. The easiest counterexample is when $A_\mu^a = 0$; $F_{\mu\nu}^a = 0$ and thus $(\delta^{ab} - n^a n^b) F^b_{\mu\nu} = 0$ holds, but $D_\mu n^a = \partial_\mu n^a \neq 0$ in general. See also Ref.~\cite{Deser:1976wj} for the necessarily condition to have physically distinct gauge potential for given field strength.}.
Only in this regime do the extra two terms in Eq.\,\eqref{eq:helicity_ir_georgi_glashow} vanish, yielding the purely topological result in Eq.\,\eqref{eq:helicity_ir_conjectured}. Consequently, the conjectured relation holds if $D_\mu n^a = 0$, i.e., in the strict IR where the perpendicular gauge fluctuations have decoupled.

Let us now summarize how to compute the helicity change in the Georgi--Glashow model, given a configuration of the Higgs field $\Phi^a$ and gauge field $A^a_\mu$ on a spacetime manifold $M$. As emphasized above, the 't Hooft tensor $\mathcal{F}_{\mu\nu}$ is only well-defined where the Higgs field does not vanish. Thus, we must excise from $M$ all regions where $\Phi^a = 0$, analogous to removing neighborhoods of 't Hooft loops in the Abelian case. The resulting punctured manifold, which we denote $M'$, may have excised regions corresponding to monopole worldlines as well as other loci where the Higgs field vanishes.

On $M'$, the 't Hooft tensor $\mathcal{F}_{\mu\nu}$ is globally defined. To compute the helicity change, we must also specify the Dirac string worldsheets. This can be done in several ways: for example, by identifying nontrivial 2-cycles in $M'$ with nonzero first Chern number of $\mathcal{F}_{\mu\nu}$ and spanning a thin Dirac sheet along each such cycle; or by globally ``combing'' the Higgs field (i.e., choosing a continuous gauge transformation $U$ that aligns $n^a$ everywhere), and identifying the loci where $U$ is discontinuous as the Dirac string worldsheets.

Once these are specified, we are ready to compute the helicity change. In ``simple'' cases where the Higgs field vanishes only on monopole worldlines,
the helicity change is given by Eq.\,\eqref{eq:helicity_change} or, more generally, Eq.\,\eqref{eq:helicity_change_generic}. In more complicated situations, we need to add other boundary integrals. To see this, it is instructive
to go back to the derivation of Eq.\,\eqref{eq:helicity_change} in Sec.\,\ref{sec:abelian_helicity_change}. The point is that in a specetime region $M''$ where the vector potential $\mathcal{A}_\mu$ is well-defined, the helicity change is given by
\begin{align}
    \mathcal{H}_\text{future} - \mathcal{H}_\text{past}
    = \int_{M''} \mathcal{F} \wedge \mathcal{F} + \int_{N} \mathcal{A} \wedge \mathcal{F},
    \label{eq:helicity_change_georgi_glashow}
\end{align}
where $\partial M'' = -N \cup \text{(Future hypersurface)} \cup -\text{(Past hypersurface)}$.
For the generic Higgs field configuration, we need to use this formula instead. 

Alternatively, one can compute the helicity directly on the past and future time slices using the UV data, as in Eq.\,\eqref{eq:helicity_ir_georgi_glashow}. The difference between these values gives the helicity change, which must agree with the result from Eq.\,\eqref{eq:helicity_change}. In the special case where the Higgs field is covariantly constant ($D_\mu n^a = 0$) on the initial and final slices, the helicity can be expressed in terms of the Chern--Simons number and the Hopf invariant, as discussed in Ref.\,\cite{Hamada:2025cwu}. In this regime, the contribution $\int_{M'} \mathcal{F} \wedge \mathcal{F}$ corresponds to the change in Chern--Simons number, $\int_M F^a \wedge F^a$, while the Dirac sheet contribution $\int_{\Sigma} F$ matches the change in the Hopf invariant.

Indeed, we may explicitly see $\int_{\Sigma} F$, or more generally $\int_{N} \mathcal{A}\wedge\mathcal{F}$, contains the change in the Hopf invariant $N_H$ when the UV degrees perpendicular to the IR ones decouple. Using the definition of the Hopf invariant, Eq.\,\eqref{eq:hopf_invariant}, the change in $N_H$ between the past and future time slices, $\Delta N_H$, satisfies 
\begin{align}
    \Delta N_H - \frac{g^2}{4\pi^2}\int_{N} \omega \wedge d\omega = \frac{g^2}{4\pi^2} \int_{M''} d(\omega \wedge d\omega) = 0,
\end{align}
where the last equality follows from the fact that $d(\omega \wedge d\omega) = d\omega \wedge d\omega$ antisymmetrizes four $d n^a$. Here, the integration must be performed not on $M'$ but on $M''$. This is because $\omega$ is not well-defined globally on $M'$; as we have discussed above, we need to ``comb the hedgehog'' to define $\omega$ globally, which introduces Dirac string worldsheets.
On the other hand, from Eq.\,\eqref{eq:helicity_ir_georgi_glashow}, 
\begin{align}
    \int_{N} \mathcal{A}\wedge\mathcal{F} = \int_{N} \qty(\mathcal{K} + 4\omega\wedge d\omega)
\end{align}
follows. Therefore, $\int_{N} \mathcal{A}\wedge\mathcal{F}$ contains the change in the Hopf invariant.

It should be noted, however, that the precise separation between these contributions can depend on the choice of excision radius around monopole worldlines: if the excised region is too small, UV degrees of freedom may affect the Chern--Simons number; if too large, the omitted region $M \setminus M'$ may contribute non-negligibly to $\int_M F^a \wedge F^a$. Care must therefore be taken in interpreting the helicity change in terms of UV and IR quantities. 

\section{Helicity change in the electroweak like theory}
\label{sec:standard_model_like}
Finally, we extend the discussion to the electroweak like theory, which is a non-Abelian gauge theory with a fundamental Higgs field. The UV gauge group is $\text{SU}(2)_L \times \text{U}(1)_Y$, and the Higgs field, $H$, is in the fundamental representation of $\text{SU}(2)_L$ with a hypercharge $Y = 1/2$. The Higgs field acquires a VEV in the vacuum,
\begin{align}
    \langle H \rangle = \begin{pmatrix} 0 \\ v/\sqrt{2} \end{pmatrix}, \label{eq:higgs_vev}
\end{align}
which breaks the gauge symmetry down to $\text{U}(1)_\text{EM}$. 
Although the theory does not have a 't Hooft--Polyakov monopole solution, again, the vector potential $\mathcal{A}_\mu$ in the low-energy $\text{U}(1)_\text{EM}$ gauge theory is not {\it a priori} well-defined if the Higgs field is not constant in spacetime. We first need to define the gauge-invariant electromagnetic field strength $\mathcal{F}_{\mu\nu}$ and then derive the vector potential $\mathcal{A}_\mu$ {\it a posteriori}.

Let us perform the same procedure as in the Georgi--Glashow model. First, we define a unit vector $n^a$. As the Higgs field is in the fundamental representation of $\text{SU}(2)_L$, we can define the unit vector as
\begin{align}
    \label{eq:hopf_projection}
    &n^a = \frac{H^\dagger \sigma^a H}{|H|^2} = h^\dagger \sigma^a h,
    &&n^a n^a = 1,
\end{align}
where $\sigma^a$ are the Pauli matrices and $h$ is the normalized Higgs field, defined as $h \equiv H / \sqrt{|H|^2}$. $n^a$ obeys the adjoint representation of $\text{SU}(2)_L$ with vanishing hypercharge. 

In the vacuum, where the Higgs field is constant, the definition of the electromagnetic gauge field $A_\mu$ and the $Z$ boson gauge field $Z_\mu$ is, using the $\text{SU}(2)_L$ gauge field $W_\mu^a$ and the hypercharge gauge field $B_\mu$,
\begin{align}
    &A_\mu = W_\mu^3 \sin \theta_W + B_\mu \cos \theta_W, 
    &&Z_\mu = W_\mu^3 \cos \theta_W - B_\mu \sin \theta_W,
\end{align}
where $\theta_W$ is the weak mixing 
angle, defined as
\begin{align}
    &\sin \theta_W = \frac{g'}{\sqrt{g^2 + g'^2}}, 
    &&\cos \theta_W = \frac{g}{\sqrt{g^2 + g'^2}},
\end{align}
where $g$ and $g'$ are the gauge couplings of $\text{SU}(2)_L$ and $\text{U}(1)_Y$, respectively.
The electromagnetic field strength is then defined as
\begin{align}
    F_{\mu\nu} = \partial_\mu A_\nu - \partial_\nu A_\mu = \qty(\partial_\mu W_\nu^3 - \partial_\nu W_\mu^3) \sin \theta_W + F^{B}_{\mu\nu} \cos \theta_W,
\end{align}
where $F^B_{\mu\nu}$ is the field strength of the hypercharge gauge field $B_\mu$.
Just like the Georgi--Glashow model~\eqref{eq:thooft_tensor}, we can write the combination $\partial_\mu W_\nu^3 - \partial_\nu W_\mu^3$ as the following gauge-invariant quantity:
\begin{align}
    \mathcal{F}^{W}_{\mu\nu} = F^{Wa}_{\mu\nu} n^a - \frac{1}{g} \epsilon^{abc} n^a D_\mu n^b D_\nu n^c,
\end{align}
where $F^{Wa}_{\mu\nu}$ is the field strength of the $\text{SU}(2)_L$ gauge field. Again, this is a closed two-form, and it reduces to $\partial_\mu W_\nu^3 - \partial_\nu W_\mu^3$ with Eq.\,\eqref{eq:higgs_vev}. For the same reason as in the Georgi--Glashow model, we use such closed two-form to define the gauge-invariant electromagnetic field strength $\mathcal{F}_{\mu\nu}$:
\begin{align}
    \mathcal{F}_{\mu\nu} = -\mathcal{F}^{W}_{\mu\nu} \sin \theta_W + F^B_{\mu\nu} \cos \theta_W,
\end{align}
where the minus sign is due to the definition of $n^a$, which is defined so that $n^3 = -1$ for Eq.\,\eqref{eq:higgs_vev}. 
We may now \emph{derive} the vector potential $\mathcal{A}_\mu$, as $\mathcal{F}_{\mu\nu}$ is closed. Equivalently, we may derive a $\text{SU}(2)_L$ invariant ``$W^3_\mu$'' field $\mathcal{W}^3_\mu$ such that $\mathcal{F}^W_{\mu\nu} = \partial_\mu \mathcal{W}^3_\nu - \partial_\nu \mathcal{W}^3_\mu$. Then, the electromagnetic vector potential and the $Z$ boson vector potential is given by
\begin{align}
    &\mathcal{A}_\mu = -\mathcal{W}^3_\mu \sin \theta_W + B_\mu \cos \theta_W, 
    &&\mathcal{Z}_\mu = -\mathcal{W}^3_\mu \cos \theta_W - B_\mu \sin \theta_W 
\end{align}
up to the $\text{U}(1)$ gauge transformation.

Just as the Georgi--Glashow model~\eqref{eq:A_F_georgi_glashow}, the field strength $\mathcal{F}^W_{\mu\nu}$ can be written as
\begin{align}
    \mathcal{F}^W = d\qty(n^a W^a) - 2 d \omega =d\qty(\mathcal{W}^3_\mu dx^\mu).
\end{align}
For the Georgi--Glashow model, we can comb the adjoint Higgs field $n^a$ globally to obtain the explicit form of $\omega$. For the electroweak-like theory, we can write $\omega$ explicitly using the fundamental Higgs field $H$.
Using Eq.\,\eqref{eq:hopf_projection},
\begin{align}
    &\epsilon^{abc} n^a(x) \partial_\mu n^b(x) \partial_\nu n^c(x)\\
    &\quad= \partial^y_\mu \partial^z_\nu \left.\qty(\epsilon^{abc} n^a(x) n^b(y) n^c(z) )\right|_{y, z \to x} \\
    &\quad= \partial^y_\mu \partial^z_\nu \left.\qty(-\frac{i}{2} h^\dagger [\sigma^b, \sigma^c] h n^b(y) n^c(z))\right|_{y, z \to x} \\
    &\quad= -2i\partial^y_\mu \partial^z_\nu \left.\qty[h^\dagger(x) h(y) h^\dagger(y) h(z) h^\dagger(z) h(x) - (y\leftrightarrow z)]\right|_{y, z \to x} \\
    &\quad= -2i \qty[\partial_\mu \qty(h^\dagger \partial_\nu h) - \partial_\nu \qty(h^\dagger \partial_\mu h)].
\end{align}
Here, we have also used
\begin{align}
    (X_1 \sigma^a X_2) (X_3 \sigma^a X_4) = 2 (X_1 X_4)(X_3 X_2) - (X_1 X_2)(X_3 X_4),
\end{align}
for arbitrary two-component vectors $X_{1,2,3,4}$.
Therefore, we can write $\omega$ as
\begin{align}
    \omega = -\frac{i}{g} h^\dagger \partial_\mu h \,dx^\mu,
\end{align}
and thereby $\mathcal{W}^3_\mu$ as
\begin{align}
    \mathcal{W}^3_\mu = n^a W_\mu^a + \frac{2i}{g} h^\dagger \partial_\mu h, \label{eq:w3_electroweak}
\end{align}
up to a closed form. Using this, we can write the vector potentials on where the Higgs field is non-vanishing.
In particular, the $Z$ boson field can be simplified as
\begin{align}
    \mathcal{Z}_\mu = -\frac{2i}{\sqrt{g^2 + g'^2}} h^\dagger D_\mu h. \label{eq:z_boson_vector_potential}
\end{align}
Note that, again, the integration (cf. Eq.~\eqref{eq:hopf_invariant})
\begin{align}
    H_n \equiv \frac{g^2}{4\pi^2} \int \omega \wedge d \omega
\end{align}
over some time-slice being homomorphic to $S^3$ is the Hopf invariant, the element of the third homotopy group $\pi_3(S^2) = \mathbf{Z}$.
On the other hand, using the explicit form of $\omega$, we can write the Hopf invariant as
\begin{align}
    H_n = N_H = -\frac{1}{4 \pi^2} \int \qty(h^\dagger d h) \wedge \qty(d h^\dagger \wedge d h),
\end{align}
which is the winding number of the Higgs field and corresponds to the third homotopy group $\pi_3(\text{SU}(2)_L \times \text{U}(1)_Y / \text{U}(1)_\text{EM}) = \pi_3(S^3) = \mathbf{Z}$. 
Mathematically, the map, Eq.\,\eqref{eq:hopf_projection},
is the projection in the Hopf fibration $S^1 \to S^3 \overset{n}{\to} S^2$ and the isomorphism between the two homotopy groups, $\pi_3(S^3) \cong \pi_3(S^2)$, is given by the pushforward of the projection, $\pi_3(S^3) \to \pi_3(S^2)$\,\footnote{This follows from the exact sequence of homotopy groups for the fibration: $\cdots \to \pi_3(S^1) \to \pi_3(S^3) \to \pi_3(S^2) \to \pi_2(S^1) \to \cdots$. Since $\pi_3(S^1) = \pi_2(S^1) = 0$, we have $\pi_3(S^3) \cong \pi_3(S^2)$.}.

We can now express the helicity in the low-energy $\text{U}(1)_\text{EM}$ gauge theory in terms of UV gauge-theoretic quantities. In general, there is no simple universal relation as in the Georgi--Glashow model, but we may ask under what conditions a relation of the type conjectured in Ref.\,\cite{Hamada:2025cwu} holds. In the Georgi--Glashow model, the simple relation, Eq.\,\eqref{eq:helicity_ir_conjectured}, is valid when the heavy gauge modes decouple from the low-energy $\text{U}(1)$ sector.

For the electroweak-like theory, the decoupling of heavy gauge modes requires that both the charged $\text{SU}(2)_L$ gauge bosons ($W^\pm$) and the $Z$ boson are absent from the low-energy dynamics. The condition for the decoupling of $W^\pm$ is the same as in the Georgi--Glashow model: $D_\mu n^a = 0$, which ensures that $(\delta^{ab} - n^a n^b) F^{Wb}_{\mu\nu} = 0$ as well.
As we have shown in Eq.\,\eqref{eq:z_boson_vector_potential}, the $Z$ boson decouples when $D_\mu h = 0$. 
Notably, $D_\mu h = 0$ implies $D_\mu n^a = 0$, so the decoupling of the $Z$ boson automatically ensures the decoupling of $W^\pm$ as well.

It is important to clarify the relationship between the decoupling of $W^\pm$ and $Z$ boson fields. One might expect that since the $Z$ boson is heavier than the $W^\pm$ (especially for $g' \gg g$), the decoupling of $W^\pm$ would automatically imply the decoupling of the $Z$ boson. However, this is not the case. The equations of motion show that $W^\pm$ fields can source the $Z$ boson, but not vice versa. Therefore, if $W^\pm$ vanishes ($D_\mu n^a = 0$), the $Z$ boson field ($D_\mu h$) does not necessarily vanish. In contrast, if the $Z$ boson vanishes ($D_\mu h = 0$), then $W^\pm$ must also vanish ($D_\mu n^a = 0$). For example, the $Z$-string solution~\cite{Nambu:1977ag,Vachaspati:1992fi} features a nonzero $Z$ boson field with vanishing $W^\pm$. The gauge field energy density, $V \sim g^2 v^2 W_\mu^+ W^{-\mu} + \frac{(g^2 + g'^2) v^2}{2} \mathcal{Z}_\mu \mathcal{Z}^\mu$, further illustrates that even if $g' \gg g$ and the $Z$ boson is very heavy, the $Z$ field $\mathcal{Z}_\mu$ need not be exactly zero, but is typically suppressed by $g/g'$ relative to the $W^\pm$ field. 

With the above condition, $D_\mu h = 0$, $\mathcal{Z} = 0$ and 
\begin{align}
\mathcal{A} = \frac{B}{\cos \theta_W} = -\frac{\mathcal{W}^3}{\sin \theta_W}.
\end{align}
Therefore, we can write the IR $\text{U}(1)_\text{EM}$ helicity,
\begin{align}
    \mathcal{H}_\text{EM} = \int \mathcal{A} \wedge \mathcal{F},
\end{align}
both in terms of the quantities in the UV $\text{U}(1)_Y$ gauge theory and in the UV $\text{SU}(2)_L$ gauge theory.
The former is simple:
\begin{align}
    e^2\mathcal{H}_\text{EM} = g'^2 \int B \wedge F^B = g'^2 \mathcal{H}_B,
\end{align}
where $e=g g'/\sqrt{g^2+g'^2}$, and $\mathcal{H}_B$ is the helicity in the UV $\text{U}(1)_Y$ gauge theory. Namely, the IR electromagnetic helicity is proportional to the UV hypercharge helicity.
On the other hand, for the latter, we may write
\begin{align}
    e^2 \mathcal{H}_\text{EM} = g^2 \int \mathcal{W}^3 \wedge \mathcal{F}^W.
\end{align}
As in the Georgi--Glashow model,  
we can reduce the integral further to (cf. Eq.\,\eqref{eq:helicity_ir_georgi_glashow})
\begin{align}
    \int \mathcal{W}^3 \wedge \mathcal{F}^W &= \int \qty( \mathcal{K} + 4 \omega \wedge d\omega),
\end{align}
since
the decoupling condition, $D_\mu n^a = 0$,
follows from $D_\mu h=0$.
Therefore, we have
\begin{align}
    \label{eq:cond_hmu}
    e^2 \mathcal{H}_\text{EM} = g'^2 \mathcal{H}_B = 16\pi^2 \qty(N_\text{CS} + N_H).
\end{align}
The second equality is the relation conjectured in Ref.\,\cite{Hamada:2025cwu} for the electroweak-like theory; we have shown that the relation holds when the massive gauge bosons, $W^\pm$ and $Z$, decouple from the low-energy $\text{U}(1)_\text{EM}$ gauge theory.

Note that $N_H$ is invariant under the variation of the fields as it is the  Maurer--Cartan integral~\cite{Weinberg:1996kr}.
Consequently, the value of $N_H$ is always an integer, and changes only when $h$ is not well-defined (which corresponds to the point where Higgs field takes zero). On the other hand, $N_\text{CS}$ can be a non-integer when the background is not vacuum. Since it costs an energy to realize Higgs zero, at low energy, the change of $\text U(1)_\text{EM} $ helicity is caused by the change of $N_\text{CS}$. This is consistent with the expectation that the helicity change of the photon does not generate the baryon asymmetry (see also the discussion in Sec.~\ref{sec:conclusion}). The similar remark also applies to the Georgi--Glashow model discussed in Sec.~\ref{sec:non_abelian_adjoint_Higgs}.

Similarly, the relation between the helicity change and the integral of the electromagnetic field strength $\mathcal{F} \wedge \mathcal{F}$ can be discussed in the same way as the Georgi--Glashow model. We first scoop out the region around the Higgs field vanishing, and then define the vector potential $\mathcal{A}_\mu$ on the remaining spacetime manifold $M'$. However, one important difference is that there is no magnetic monopole solution in the electroweak like theory and hence no 't Hooft loop representing (not too heavy) dynamical monopole worldline. Therefore, we do not need to introduce the Dirac string worldsheet $\Sigma_i$ in this case. In other words, the vector potential $\mathcal{A}_\mu$ can be defined globally on the whole $M'$, as Eq.\,\eqref{eq:w3_electroweak} shows.

Although the electromagnetic vector potential $\mathcal{A}_\mu$ can be defined globally on $M'$, the helicity change contains other contributions than the integral of $\mathcal{F} \wedge \mathcal{F}$; the helicity change formula is given as 
\begin{align}
    \mathcal{H}_\text{EM,future} - \mathcal{H}_\text{EM,past} = \int_{M'} \mathcal{F} \wedge \mathcal{F} + \int_{N} \mathcal{A} \wedge \mathcal{F},
\end{align}
where $N$ is the boundary of $M'$ excluding the initial and final time slices and thus given by the boundary of the region where the Higgs field is vanishing\,\footnote{The direction of $N$ is the same as Eq.\,\eqref{eq:helicity_change_georgi_glashow}.}. Any contribution from the ``Nambu monopole--anti-monopole pair''-like configuration\,\cite{Vachaspati:2001nb,Hamada:2025cwu} should be included in this boundary integral.

We may also express the helicity change in terms of the UV gauge-theoretic quantities. As in the Georgi--Glashow model, if the Higgs field is covariantly constant on the initial and final time slices, $D_\mu h = 0$, the helicity on each time slice can be expressed in terms of the $\text{SU}(2)_L$ Chern--Simons number and the Higgs winding number. Again, it is tempting to identify the integral $\int_{M'} \mathcal{F} \wedge \mathcal{F}$ with the change in the Chern--Simons number, $\int_M F^a \wedge F^a$, and the boundary integral $\int_{N} \mathcal{A} \wedge \mathcal{F}$ with the change in the Higgs winding number. However, as in the Georgi--Glashow model, care must be taken in interpreting these contributions, as they can depend on the choice of excision radius around regions where the Higgs field vanishes.

\section{Conclusion and Discussion}
\label{sec:conclusion}
In this paper, we have discussed the change of $\text{U}(1)$ magnetic helicity between two time slices. For the effective Abelian gauge theory, if there are no magnetic monopoles in the spacetime region between the two time slices, the helicity change is given by the integral of the electromagnetic field strength $F \wedge F$ over the spacetime region. However, we have found that if there are magnetic monopoles, we need to introduce the Dirac string worldsheet and add the integral of the electromagnetic field strength over the Dirac string worldsheet to the helicity change formula.

We have then extended the discussion to the case where the $\text{U}(1)$ gauge theory is embedded in a non-Abelian gauge theory with a Higgs field, which breaks the non-Abelian gauge symmetry down to $\text{U}(1)$. We have shown that the helicity change formula is similar to the effective Abelian gauge theory case. However, to define the electromagnetic vector potential, we need to excise the region where the Higgs field is vanishing. Therefore, the boundary of the excised region also contributes to the helicity change formula. 

We have also discussed the relation between the helicity in the low-energy $\text{U}(1)$ gauge theory and the UV theoretic quantities, such as the Chern--Simons number and the Higgs winding number. We have found that, as conjectured in Ref.\,\cite{Hamada:2025cwu}, the helicity can be expressed as the sum of the Chern--Simons number and the Higgs winding number when the heavy gauge bosons decouple from the low-energy $\text{U}(1)$ gauge theory. We have clarified the condition for the decoupling of the heavy gauge bosons.

An immediate application of our results is to the study of the generation of the baryon asymmetry of the universe via the chiral anomaly in the standard model.
The baryon number symmetry in the standard model is anomalous under the $\text{SU}(2)_L \times \text{U}(1)_Y$ gauge symmetry, as 
\begin{align}
    \partial_\mu j_B^\mu = \frac{3g^2}{32 \pi^2} F^{Wa}_{\mu\nu} \tilde{F}^{Wa\mu\nu} - \frac{3g'^2}{32 \pi^2} F^B_{\mu\nu} \tilde{F}^{B\mu\nu}.
\end{align}
Therefore, the change in the baryon number, $\Delta Q_B$, is related to the change in the Chern--Simons number of the $\text{SU}(2)_L$ gauge field and the helicity of the hypercharge gauge field, $\Delta Q_B = 3 \left(\Delta N_\text{CS} - g'^2 \Delta \mathcal{H}_B/16\pi^2\right)$.
On the other hand, $\text{U}(1)_\text{EM}$ helicity $\mathcal{H}_\text{EM}$ today satisfies $e^2 \mathcal{H}_\text{EM} = g'^2 \mathcal{H}_B = 16\pi^2 \qty(N_\text{CS} + N_H)$ from Eq.\,\eqref{eq:cond_hmu}. In previous literature~\cite{Kamada:2016cnb}, 
it has been argued that baryon asymmetry is induced in the system at the electroweak symmetry breaking (EWSB) if helical hypermagnetic fields are present beforehand. 
The argument assumes that the helicity of the hypercharge gauge field before EWSB and that of the electromagnetic field after EWSB are of the same order, such that $\mathcal{H}_B^\mathrm{before} \sim \mathcal{H}_\text{EM}^\mathrm{after} \sim \Delta N_\text{CS} \sim - \Delta \mathcal{H}_B$. 
Although this setup differs from the present one and our discussion cannot be directly applied, it should be noted that in Ref.~\cite{Kamada:2016cnb} the contribution from the Higgs winding number has been neglected.
The contribution of the Higgs winding number may not be negligible in general. With the nonzero Higgs winding number change, $\Delta Q_B \propto (\Delta N_\mathrm{CS} - g'^2 \Delta \mathcal{H}_B/16\pi^2)$ could be vanishing even if 
$\mathcal{H}_B^\mathrm{before}$ and $\mathcal{H}_\text{EM}^\mathrm{after}$
are nonzero although in general we also take the fermion number on the boundary, e.g., the calculation similar to App.\,\ref{app:dyon_fermion_number}, into consideration, which could well compensate the Higgs winding numbers.
Our observation implies that the baryon asymmetry in Ref.~\cite{Kamada:2016cnb} might have been overestimated.
See also Ref.\,\cite{Hamada:2025ooy} for a potential modification of the constraints on the primordial magnetic fields from the baryon asymmetry.
We need more careful, possibly numerical, studies to estimate the baryon number generation in these scenarios.

\acknowledgments
This work was supported by JSPS KAKENHI Grant Nos.\ 24K17042 [HF], 25H00638 [HF], JP24H00976 [YH], JP24K07035 [YH], JP24KF0167 [YH], JP23K17687 [KK], by MEXT Leading Initiative for Excellent Young Researchers Grant No.JPMXS0320210099 [YH], and by the National Natural Science Foundation of China (NSFC) under Grant No. 12347103 [KK].
In this research work, HF used the UTokyo Azure (\url{https://utelecon.adm.u-tokyo.ac.jp/en/research_computing/utokyo_azure/}). 
HF would like to thank Thanaporn Sichanugrist for usuful discussions. HF and KK would like to thank Mikhail Shaposhnikov and Tanmay Vachaspati for useful discussions.

\appendix
\section{Notation}
\label{app:notations}
We take the spacetime metric signeture to be $(+, -, -, -)$.
For the totally antisymmetric tensor $\epsilon^{\mu\nu\rho\sigma}$, we take $\epsilon^{0123}=+1$.
The four-vector potential is defined as
\begin{align}
    A^\mu = (\varphi, \vec{A}),
\end{align}
where $\varphi$ is the scalar potential and $\vec{A}$ is the vector potential. Therefore,
\begin{align*}
    A_\mu = (\varphi, -\vec{A}).
\end{align*}
Note that this is the same convention as Ref.\,\cite{Peskin:1995ev}.
The electromagnetic fields are defined as
\begin{align}
    &\vec{E} = -\nabla \varphi - \frac{\partial \vec{A}}{\partial t}, 
    &&\vec{B} = \nabla \times \vec{A}.
\end{align}
The electromagnetic field strength tensor is given by
\begin{align}
    \label{eq:field_strength_tensor}
    F_{\mu\nu} &= \partial_\mu A_\nu - \partial_\nu A_\mu 
    = \begin{pmatrix}
        0 & E_x & E_y & E_z \\
        -E_x & 0 & -B_z & B_y \\
        -E_y & B_z & 0 & -B_x \\
        -E_z & -B_y & B_x & 0
    \end{pmatrix}.
\end{align}
Namely, we have
\begin{align}
    &F_{0i} = E_i, 
    &&F_{ij} = -\epsilon^{ijk} B_k, 
    \label{eq:electromagnetic_field}
\end{align}
where $i, j, k$ run over the spatial indices $1, 2, 3$ (or $x, y, z$), and $\epsilon^{ijk}$ is the Levi--Civita symbol in three dimensions, defined as $\epsilon^{123} = 1$.

The electromagnetic field strength can also be expressed in terms of the differential form notation:
\begin{align}
    A &\equiv A_\mu dx^\mu \\
    F &\equiv dA = \frac12 F_{\mu\nu}dx^\mu \wedge dx^\nu.
\end{align}
The wedge product of two field strength tensors is given by
\begin{align}
    F \wedge F &= \frac{1}{4} F_{\mu\nu} F_{\rho\sigma} dx^\mu \wedge dx^\nu \wedge dx^\rho \wedge dx^\sigma \\
    &= \frac{1}{4} \qty(F_{01} F_{23} \times 8 + F_{02} F_{13} \times 8 + F_{03} F_{12} \times 8) dV \\
    &= 2 \qty(-E_x B_x - E_y B_y - E_z B_z) dV \\
    &= -2 \vec{E} \cdot \vec{B} dV,
\end{align}
where $dV = dx^0 \wedge dx^1 \wedge dx^2 \wedge dx^3$ is the four-dimensional spacetime volume element. The numerical factor of $8$ comes from the number of distinct permutations of the indices in each term: $2$ from exchanging $0 \leftrightarrow 1$, $2$ from $2 \leftrightarrow 3$, and $2$ from swapping the pairs $(0,1) \leftrightarrow (2,3)$, giving $2 \times 2 \times 2 = 8$.

We also note the Lorentz transformation properties of the electromagnetic fields. Suppose we have a Lorentz boost in the $z$ direction with velocity $v$, which is non-relativistic, i.e., $v \ll 1$. The Lorentz transformation is given by
\begin{align}
    \vec{E}' &\simeq \vec{E} + \vec{v} \times \vec{B}, \label{eq:Lorentz_transformation_E}\\
    \vec{B}' &\simeq \vec{B} - \vec{v} \times \vec{E},
\end{align}
where $\vec{E}'$ and $\vec{B}'$ are the electric and magnetic fields in the boosted frame, respectively. We have dropped the terms of order $v^2$ and higher.

The non-Abelian gauge transformation of the vector potential is given by
\begin{align}
    A_\mu \to \frac{i}{g} U(x) \partial_\mu U^\dagger(x) + U(x) A_\mu U^\dagger(x),
    \label{eq:non-abelian_gauge_transformation}
\end{align}
where $U(x)$ is a local gauge transformation and $g$ is the gauge coupling constant. Here, the vector potential $A_\mu$ takes the matrix form, i.e., $A_\mu = A_\mu^a T^a$, where $T^a$ are the generators of the gauge group in the fundamental representation.

The covariant derivative of a field $\phi$ is defined as
\begin{align}
    D_\mu \phi = \partial_\mu \phi - i g A^a_\mu T_R^a \phi,
\end{align}
where $T_R^a$ are the generators of the gauge group in the representation $R$ of the field $\phi$. In particular, for the adjoint representation, the covariant derivative is given by
\begin{align}
    \label{eq:covariant_derivative}
    D_\mu \phi^a = \partial_\mu \phi^a + g f^{abc} A_\mu^b \phi^c,
\end{align}
where $f^{abc}$ are the structure constants of the gauge group.

We may also use the differential form notation for the non-Abelian gauge theory. The vector potential and field strength are defined as
\begin{align}
    \label{eq:non-abelian_field_strength}
    A^a &= A_\mu^a dx^\mu, \\
    F^a &= dA^a + \frac{1}{2} g f^{abc} A^b \wedge A^c = \frac{1}{2} F_{\mu\nu}^a dx^\mu \wedge dx^\nu.
\end{align}
The wedge product of two field strength tensors is given by
\begin{align}
    F^a \wedge F^a = d\mathcal{K},
\end{align}
where the Chern--Simons form $\mathcal{K}$ is defined as
\begin{align}
    \label{eq:non-abelian_cs}
    \mathcal{K} = A^a \wedge d A^a + \frac{1}{3} g f^{abc} A^a \wedge A^b \wedge A^c.
\end{align}

\section{Change of the fermion chiral charge under the presence of dyon}
\label{app:dyon_fermion_number}
In this appendix, we show that the change of the fermion chiral charge under the presence of the dyon is also well-defined, as in the case of the magnetic helicity. For simplicity, let us consider a manifold $M$ which only has the boundary at the past and final future slices, $\partial M = (-\mathcal{S}_\text{past}) \cup \mathcal{S}_\text{future}$ and there is only one 't Hooft--Wilson loop, which represents the worldline of the dyon. For simplicity, we assume that the dyon worldline is $S^1$ with the length $L$.
We scoop out the closed region around the dyon worldline, and denote the remaining manifold as $M'$. 
We also assume that the dyon enjoys the unit electric and magnetic charges. 

Suppose we have one Dirac fermion $\psi$ with the unit electric charge, which couples to the $\text{U}(1)$ gauge field $A_\mu$. The axial current $3$-form is defined as
\begin{align}
    j_5 = -\frac{1}{6} \epsilon_{\mu\nu\rho\sigma} j_5^\mu dx^\nu \wedge dx^\rho \wedge dx^\sigma,
\end{align}
where
\begin{align}
    j_5^\mu = \bar{\psi} \gamma_5 \gamma^\mu \psi.
\end{align}
The minus sign in the definition of $j_5$ is due to the convention of the Levi--Civita symbol in four dimensions, $\epsilon_{0123} = -1$.
The axial anomaly equation is given by
\begin{align}
    d j_5 = \partial_\mu j_5^\mu dV = \frac{e^2}{16\pi^2} \epsilon^{\mu\nu\rho\sigma} F_{\mu\nu} F_{\rho\sigma} dV = \frac{e^2}{4\pi^2} F \wedge F.
\end{align}

Let the radius of the excised region around the dyon worldline be $\epsilon$. The scooped out region is $S^1\times B_\epsilon^3$, where $B_\epsilon^3$ is the $3$-dimensional ball with radius $\epsilon$.
We show that the change of the fermion chiral charge is well-defined in the limit $\epsilon \to 0$.
The change of the fermion chiral charge is given by
\begin{align}
    Q_5^\text{future} - Q_5^\text{past} &= \int_{M'} d j_5 - \int_{\partial\qty(S^1 \times B_\epsilon^3)} j_5 \label{eq:fermion_change_dyon}
\end{align}
where $Q_5 = \int_{\Sigma} j_5$ is the fermion chiral charge on the time slice $\Sigma$. Just as in the case of the magnetic helicity, both the first and second terms on the right-hand side are divergent in the limit $\epsilon \to 0$, but their sum is finite.
To see this, let us first evaluate the first term. Using Eq.\,\eqref{eq:fft_divergent}, we have
\begin{align}
    \int_{M'} d j_5 &= \frac{e^2}{4\pi^2} \int_{M'} F \wedge F \nonumber \\
    &= -\frac{e^2L}{4\pi^2\epsilon} + \mathcal{O} (\epsilon^0). \label{eq:fermion_change_dyon_bulk}
\end{align}

On the other hand, the second term is the fermion chiral charge on the boundary of the excised region, the three-dimensional manifold $S^1\times S^2$. There is ambiguity in defining the fermion chiral charge on this manifold since the fermion chiral charge of the vacuum is not uniquely defined. Here, for simplicity, we assume that the fermion chiral charge on the surface is vanishing when the electric charge of the dyon is zero, and we adiabatically\,\footnote{Here, we imagine introducing a fictitious time-like parameter to describe the adiabatic change of the electric charge in addition to $S^2 \times S^1$. This parameter is a purely mathematical device and is distinct from the physical time in the original spacetime manifold $M$.} increase the electric charge of the dyon from zero to one and evaluate the change of the fermion chiral charge on the surface. This can be done by the adiabatic change of the $\theta$ angle from $0$ to $2\pi$\,\cite{Witten:1979ey}.

Since the chiral charge is vanishing when the electric charge of the dyon is zero, the change of the fermion chiral charge on the surface is coming from the vacuum state of the massless mode on the surface. Let us count the number of the massless modes on the surface. Due to the magnetic flux from the dyon, the motion of the fermion on the $S^2$ is quantized into the Landau levels. The number density of the lowest Landau level is given by
\begin{align}
    n = 2\frac{eB}{2\pi} = \frac{e g_m}{4 \pi^2 \epsilon^2} = \frac{1}{2\pi \epsilon^2},
\end{align}
where $B = g_m / (4\pi \epsilon^2)$ is the magnetic field strength on the surface, and the factor $2$ comes from the particle and antiparticle contributions. The area of the $S^2$ is $4\pi \epsilon^2$, and thus the total number of the lowest Landau level is given by
\begin{align}
    N = n \times 4\pi \epsilon^2 = 2.
\end{align}
The same result can be obtained by the Atiyah--Singer index theorem; see Example 12.6 in Ref.\,\cite{Nakahara:2003nw}, for instance.
Now, we have two massless modes on the $S^1$ direction and this is equivalent to the massless $1+1$ dimensional Dirac fermion. As we adiabatically change the electric charge, the vector potential on the $S^1$ direction, $A_{S^1}$,  changes from $0$ to $\frac{e}{4\pi \epsilon}$. As $eA_{S^1}$ increases by $2\pi/L$, the vacuum state of the massless mode changes and the fermion chiral charge on the surface changes by $2$\,\cite{Peskin:1995ev}. Therefore, the change of the fermion chiral charge on the surface is given by
\begin{align}
    -\int_{\partial\qty(S^1\times B_\epsilon^3)} j_5 = \int_{S^1\times S^2} j_5= N \times \frac{e^2 L}{8\pi^2 \epsilon} = \frac{e^2 L}{4\pi^2 \epsilon}.
    \label{eq:fermion_change_dyon_surface}
\end{align}
Combining Eqs.\,\eqref{eq:fermion_change_dyon_bulk} and \eqref{eq:fermion_change_dyon_surface}, we find that the change of the fermion chiral charge, Eq.\,\eqref{eq:fermion_change_dyon}, is finite and well-defined in the limit $\epsilon \to 0$.

\bibliographystyle{JHEP}
\bibliography{papers}

\end{document}